\documentclass{article}

\usepackage{geometry}

\usepackage{times}
\usepackage{w-thm}
\usepackage[authoryear]{natbib}
\usepackage{url}
\usepackage{amsmath}
\usepackage{amsfonts}
\usepackage{multirow}
\usepackage{makecell}

\newcommand{\E}{{\mathbb{E}}}
\newcommand{\Cov}{{\operatorname{Cov}}}
\newcommand{\R}{{\mathbb{R}}}

\theoremstyle{plain}

\theoremstyle{definition}

\usepackage[]{graphicx}
\chardef\bslash=`\\ 

\title{Functional Data Analysis:\\An Introduction and Recent Developments}

\author{
    Jan Gertheiss\thanks{Corresponding author: \texttt{jan.gertheiss@hsu-hh.de}} \\
    Department of Mathematics and Statistics \\
    School of Economics and Social Sciences \\
    Helmut Schmidt University, Hamburg, Germany
\and
    David Rügamer \\
    Department of Statistics, LMU Munich \\
    Munich Center for Machine Learning, Munich, Germany
\and
    Bernard X.W. Liew \\
    School of Sport, Rehabilitation and Exercise Sciences \\
    University of Essex, Essex, UK
\and
    Sonja Greven \\
    Chair of Statistics \\
    School of Business and Economics \\
    Humboldt-Universität zu Berlin, Berlin, Germany
}
\date{}

\begin{document}

\maketitle

\begin{abstract}
Functional data analysis (FDA) is a statistical framework that allows for the analysis of curves, images, or functions on higher dimensional domains. The goals of FDA, such as descriptive analyses, classification, and regression, are generally the same as for statistical analyses of scalar-valued or multivariate data, but FDA  brings additional challenges due to the high- and infinite dimensionality of observations and parameters, respectively. This paper provides an introduction to FDA, including a description of the most common statistical analysis techniques, their respective software implementations, and some recent developments in the field. The paper covers fundamental concepts such as descriptives and outliers, smoothing, amplitude and phase variation, and functional principal component analysis. It also discusses functional regression, statistical inference with functional data, functional classification and clustering, and machine learning approaches for functional data analysis. The methods discussed in this paper are widely applicable in fields such as medicine, biophysics, neuroscience, and chemistry, and are increasingly relevant due to the widespread use of technologies that allow for the collection of functional data. Sparse functional data methods are also relevant for longitudinal data analysis. All presented methods are demonstrated using available software in R by analyzing a data set on human motion and motor control. To facilitate the understanding of the methods, their implementation, and hands-on application, the code for these practical examples is made available on Github: \url{https://github.com/davidruegamer/FDA_tutorial}.
\end{abstract}

\maketitle                   

\section{Introduction}

In functional data analysis (FDA), curves, images, or functions on higher dimensional domains constitute the observations and the objects of interest in the analysis \citep{ramsay_functional_2005}. The goals of FDA, such as descriptive analyses, classification, regression, etc., are often the same as for statistical analyses of scalar-valued or multivariate data, and many corresponding methods have been transferred to function-valued data, while FDA also brings additional challenges including infinite/high dimensionality of observations and parameters as well as misalignment. In contrast to simpler methods that reduce the functional observations to scalar summary values, FDA retains all important information by directly using the functional observations in the analysis. Functional data occur in many different fields, including medicine \citep{sorensen_2013}, biophysics \citep{liew2020classifying}, neuroscience \citep{Ruegamer.2018} or chemistry \citep{Brockhaus.2015}. Nowadays, technologies such as imaging techniques, electroencephalographs and
 electrocardiographs,  gyroscopes, or accelerometers, standard in bio-medical research or even built into everyday devices such as smartphones, controllers, or wristwatches, allow for seamlessly collecting functional data. Consequently, many research fields observe an increased interest in methods to analyze this type of data. In this paper, we describe a range of the most common statistical analysis techniques in FDA, their respective software implementations, and some more recent developments in the field. We aim to provide an introduction and tutorial for some of the most important topics in FDA, while providing a partial review with further reading for some additional, more advanced topics. 

\subsection{Functional data, longitudinal data, and time series}

Data sets collected over time occur in different fields and under different names and we thus briefly discuss differences between FDA and longitudinal as well as time-series data for readers new to the field. 

In health and social sciences, repeated observations over time on the same subjects or observational units are often termed longitudinal data (LD) or panel data. LD usually involves only a few repeated measurements per observation unit (e.g., subject observed before medication and at two follow-up appointments) with potentially different (numbers of) time points for different observation units. While functional data (FD) with different measurement points across observations exists (so-called sparse FD), the data is often recorded on the same (time) interval, with the same frequency, and exhibits a large number of measurements per observation unit. Methods to analyze FD and LD thus traditionally differ in how data is understood and treated. While the repeated measurements of observation units in the analysis of LD are often accounted for by using random effects in a parametric model, FD usually treats the observed sequences per measurement unit as observations of a smooth process observed at discrete time points. It thus models this data using non- or semi-parametric methods, aided by the typically larger number of available measurement points per curve. Nevertheless, there is an overlap between the two areas, and ideas and methods from (sparse) functional data analysis have more recently been used to make longitudinal data models less parametric and more flexible \citep[e.g.][]{yao2005functional,goldsmith2013corrected,kohler2017flexible}.

Time-series analysis, on the other hand, typically considers a single observation of a stochastic process on an equidistant time grid from a single observation unit (e.g., values of a share at one stock market observed over time). It is thus different from LD and FD where repeated observations on each of a sample of observational units are available. Due to the limited amount of data, methods usually focus on parametric estimation and often aim for good forecasting performance for unseen future time points. Goals and data structure thus differ from LD and FDA. If a sample of different time series is available, however, as, e.g., considered in the field of time series classification, and the focus is less on parametric methods, the setting can also be considered from a functional data viewpoint.

\subsection{Notation} \label{sec:notation}

In this paper, we will usually focus on a sample of functional data $x_i$ considered to be realizations of random
functions $X_i, i= 1, \dots, n$, with values $X_i(t) \in \mathbb{R}, t \in \mathcal{T}$, over an interval $\mathcal{T} \subset \mathbb{R}$. Note that both $t$ and $X_i(t)$ can be generalized to take values in some other domain than $\mathbb{R}$, but this will not be the focus here.
The mean function will be denoted by $\E(X(t)) = \mu(t)$ and the covariance function by $\Cov(X(s),X(t)) = \Sigma(s,t)$, $s,t \in \mathcal{T}$. In practice, the functional data is only observed  on a discrete grid of measurement points $t_{il} \in \mathcal{T}$, 
giving observations $x_i(t_{il})$,
$l=1, \dots, L_i, i=1, \dots, n$. For so-called dense functional data, the grid is equal across functions and (relatively) dense in $\mathcal{T}$. For so-called sparse functional data, the grids are curve-specific and can be relatively sparse, but should together still cover $\mathcal{T}$ well. If the grids differ but are relatively dense in $\mathcal{T}$, one also speaks of irregular functional data. While a common assumption in FDA is smoothness of the functions, they are often observed with additional measurement erorr. Then, we observe
$X_i(t_{il}) = \widetilde{X}_i(t_{il}) + \varepsilon_{il}$
 at $t_{il} \in \mathcal{T}, l=1, \dots, L_i, i=1, \dots, n$, where $\widetilde{X}_i(t)$ is the underlying (smooth) function and $\varepsilon_{il}$ are independent and identically distributed (i.i.d.) noise terms. Functions are commonly considered as elements of the $\mathcal{L}^2$ space of functions
\begin{equation*}
\mathcal{L}^2(\mathcal{T}) = \left\{f: \mathcal{T} \rightarrow \R \left| \int_{\mathcal{T}} f^2(t) dt < \infty\right\}\right. \quad \text{with scalar product} \; 
< f,g> = \int_{\mathcal{T}} f(t) g(t) dt
\end{equation*}
for functions $f,g \in \mathcal{L}^2(\mathcal{T})$,
which induces the norm $\| f\|_{\mathcal{L}^2} = < f,f>^{1/2}$ and a Hilbert space structure. 

\subsection{Running data example(s): human motion and motor control}

Data from the biomechanical analysis of human motion and motor control almost always exhibit temporal, spatial, or both spatio-temporal characteristics within regular discrete bounds (e.g., gait events). For example, the flexion-extension knee joint angle during running over the stance phase of running reflects a one-dimensional (1D) time-series variable \citep{pataky2010generalized}; the plantar pressure distribution of the foot at an instant during standing reflects a 2D spatial variable \citep{montagnani2021pedobarographic}, and the 3D bone strain patterns of the femur at an instant during an exercise reflects a spatial variable \citep{martelli2014strain}. The typical approach to analyzing biomechanical data is to first reduce multi-dimensional data into a single scalar variable (e.g., maximal value), followed by statistical inference. However, statistically analyzing discretized biomechanical data could, in some instances, lead to elevated Type 1 and 2 errors \citep{pataky2015zero,pataky2016probability,robinson2015statistical}. In addition, it is commonly the interest of researchers to not only know ``what'' is statistically different but, in human movement, ``where'' and ``when'' the differences occur. One approach used to answer these questions is functional data analysis, allowing, e.g., in functional regression to model repeated measurements and to include functional biomechanical variables as the outcome \citep{warmenhoven2018force}, the covariates \citep{liew2020classifying}, or both \citep{liew.2021}.

The data we use in this paper are derived from three primary sources –- a publicly available running dataset \citep{fukuchi2017public}, and two datasets from the third author’s (BL) research on load carriage running \citep{liew2016a,liew2016b}. Details of the experimental methods will be briefly mentioned here.
In the first publicly available dataset on running (with sample size $n = 28$) in healthy adults \citep{fukuchi2017public}, running was performed on a dual-belt, force-instrumented treadmill (300~Hz; Bertec, USA),  whilst lower-limb kinematic trajectories were captured with 12 optoelectronic cameras (150~Hz; Motion Analysis Corporation, USA). Participants performed running with shoes at 2.5 m/s, 3.5 m/s, and 4.5 m/s. The second dataset comes from a previously published work investigating the effects of load carriage on running biomechanics ($n = 31$) \citep{liew2016b}, which involved participants performing overground running across in-ground embedded force platforms (2000~Hz, AMTI, Watertown, MA), while carrying three load conditions (0\%, 10\%, 20\% body weight (BW)) across three velocities (3.0 m/s, 4.0 m/s, 5.0 m/s). Lower-limb kinematic trajectories were captured using an 18 camera motion capture system (Vicon T-series, Oxford Metrics, UK; 250~Hz).
The last dataset comes from a project investigating the influence of strength training on load carriage running biomechanics ($n = 31$) \citep{liew2016a}. Participants performed overground running at a fixed velocity of 3.5 m/s ($\pm$10\%) while carrying two load conditions (0\%, 20\% BW). Motion capture equipment was identical to \citet{liew2016a}. 

For all datasets, three-dimensional (3D) bilateral lower-limb kinematics (joint angle, velocity, acceleration) and kinetics (moments) of the ankle, knee, and hip joints were extracted as variables, which yields $3 \times 4 \times 3 = 36$ variables as the combinations of the entries in the first three columns of Table~\ref{tab:vars}. All kinematic and moment variables were time normalized to 101 data points within the stance phase of each lower limb. The joint moment was normalized to body mass (N/kg).

\renewcommand{\arraystretch}{1.2}
\begin{table}[ht]
    \centering
\begin{tabular}{c|c|ccc}
    body part & kinematics/kinetics & axis & resp. & type of motion\\
    \hline\hline
    \multirow{4}{*}{\makecell{ankle \\  hip \\knee}} & acceleration $\left(\frac{\text{degree}}{\text{sec}^2}\right)$ & \multirow{4}{*}{\makecell{anterior-posterior \\ \\ medial-lateral \\ \\ vertical }} & & \multirow{4}{*}{\makecell{inversion-eversion (ankle) /\\ abduction-adduction (hip, knee) \\ plantarflexion-dorsiflexion (ankle) /\\ flexion-extension (hip, knee)\\ abduction-adduction (ankle) / \\ axial rotation (hip, knee)}}\\
     & angle ($\text{degree}$) & \\
     & moment $\left(\frac{\text{Nm}}{\text{Kg}}\right)$ & \\
     & velocity $\left(\frac{\text{degree}}{\text{sec}}\right)$& \\
\end{tabular}
    \caption{Variables available in the human motion and motor control data example, obtained as the combinations of `body part', `kinematics/kinetics', and `axis'; the latter also corresponds to specific types of motion, which may differ between body parts.}
    \label{tab:vars}
\end{table}
\renewcommand{\arraystretch}{1}

\begin{figure}[htb]
\begin{center}
\includegraphics[width=70mm]{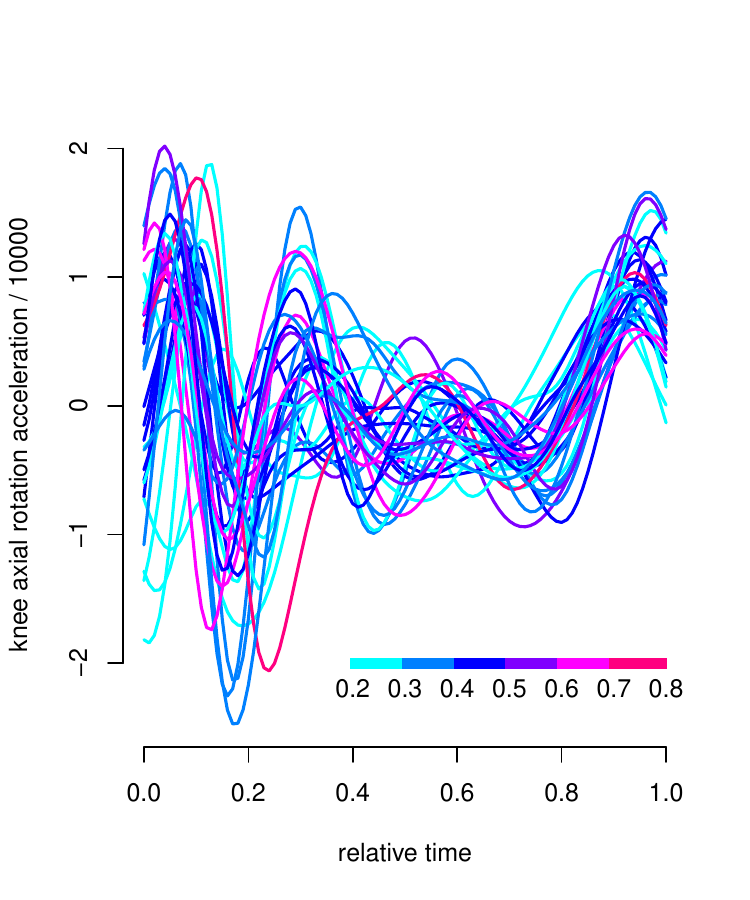}
\includegraphics[width=70mm]{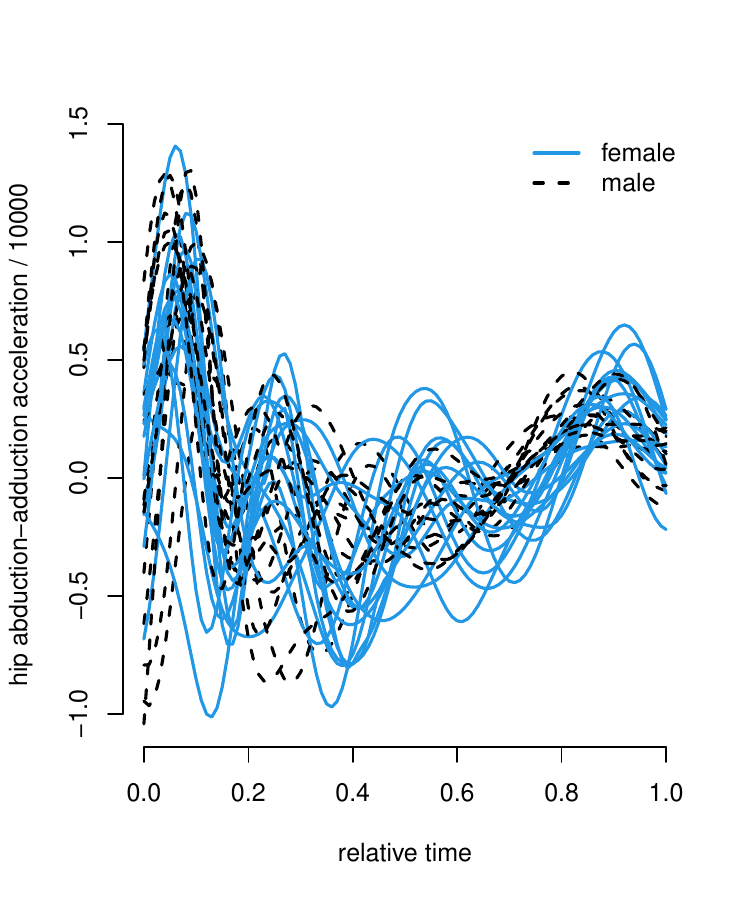}
\caption{Knee axial rotation acceleration measurements (left) with colors corresponding to the maximum moment, and hip abduction--adduction acceleration profiles (right) with colors corresponding to subjects' sex.}\label{fig:examples}
\end{center}
\end{figure}

For illustration, Figure~\ref{fig:examples}~(left) shows the knee flexion-extension acceleration (divided by 10,000) across the stance phase of running for all participants from the second study \citep{liew2016b}, when running at 3~m/s with no backpack. Colors correspond to the maximum moment. One goal of the data analysis could be to investigate how acceleration profiles relate to the often-used scalar quantity `maximum moment'. In the right panel of Figure~\ref{fig:examples}, the hip abduction-adduction acceleration is shown for females (solid blue) and males (dashed black). Here, it could be interesting to study if and where males and females differ in their profiles. 

The remainder of this paper is structured as follows. We first introduce fundamental aspects of functional data in Section~\ref{sec:fundamentals} including descriptive analysis methods, smoothing approaches, amplitude and phase variation, and functional principal component analysis. In Section~\ref{sec:funreg} we then summarize different parametric and non-parametric regression approaches before discussing methods for statistical inference with functional data in Section~\ref{sec:inference}. Methods beyond regression such as classification and clustering are considered in Section~\ref{sec:beyond} followed by an outline of machine and deep learning approaches in Section~\ref{sec:ml}. In a final outlook section (Section~\ref{sec:outlook}) we 
briefly discuss relationships of FDA to other fields, as well as FDA beyond one-dimensional functional observations. All sections are complemented with a selection of existing software packages that implement the respective approaches.

\section{Fundamentals}\label{sec:fundamentals}

\subsection{Descriptives and Outliers} \label{descriptives}

The first step in FDA usually is a descriptive analysis of the data, including visualizations and potentially the identification of outlying observations that may be influential in the analysis. For scalar data, this is relatively straightforward using tools such as a boxplot. For functional data, the task is more complex. In particular, outlying curves can differ from the majority of the sample by the range of their function values (``magnitude outliers") and/or by their shape  (``shape outliers") \citep[see][]{hyndman2010rainbow}. 

It is thus first necessary to define a notion of outlyingness or centrality of such data, which can be done using so-called functional data depths, of which different versions exist. \cite{sun2011functional} use the band-depth or modified band-depth of \cite{lopez2009concept}, which looks at the fraction of bands between pairs of other curves in the sample that contain (part of) a given curve, to define a functional boxplot. The sample median then is the curve with the highest functional depth, whereas the 50\% central region (corresponding to the inter-quartile range of a standard boxplot) is defined to envelop the 50\% most central curves in the sample with the highest depth. This region is inflated by 1.5 to generate the analog of whiskers in the boxplot and any curve lying outside this region is flagged as an outlier.

Alternative visualizations include the rainbowplot, functional bagplot, and functional highest density region boxplot of \cite{hyndman2010rainbow}, which are built on corresponding versions for multivariate data for the first two scores of a functional principal analysis (see Section \ref{sec:FPCA}), thus capturing only the two main modes of variation in the data. A simple visualization method can also be to color functions according to another variable of interest, e.g., to inform a regression analysis (compare Figure~\ref{fig:examples}, left), or according to depth. The outliergram of \cite{arribas2014shape} specifically
aims to detect shape outliers.  \cite{dai2018multivariate} discuss visualization and outlier detection for multivariate functional data.

\begin{figure}[htb]
\begin{center}
\includegraphics[width=70mm]{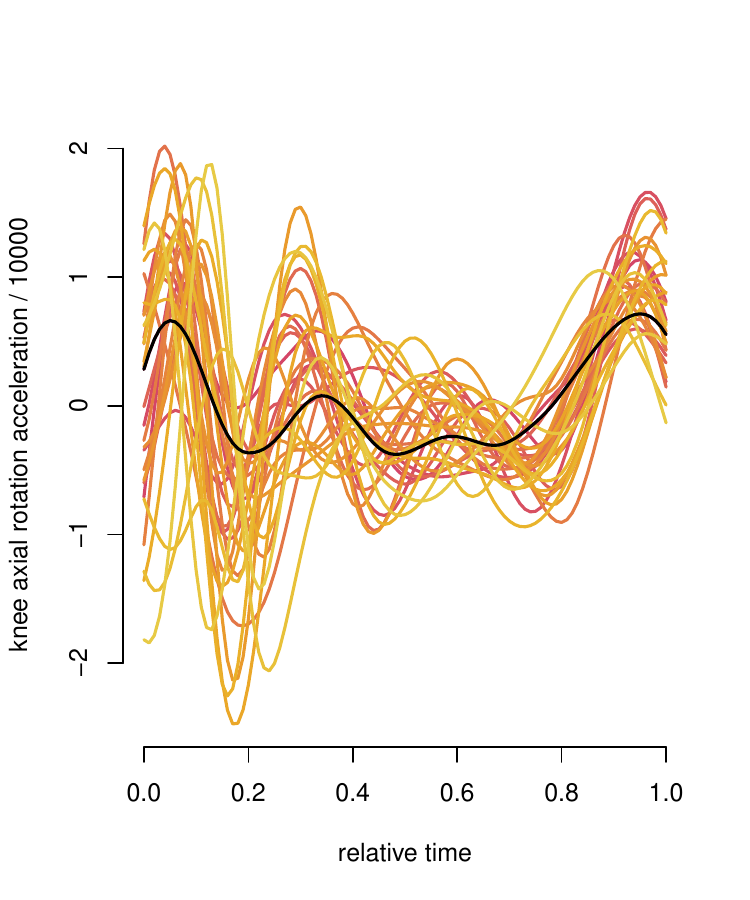}
\includegraphics[width=70mm]{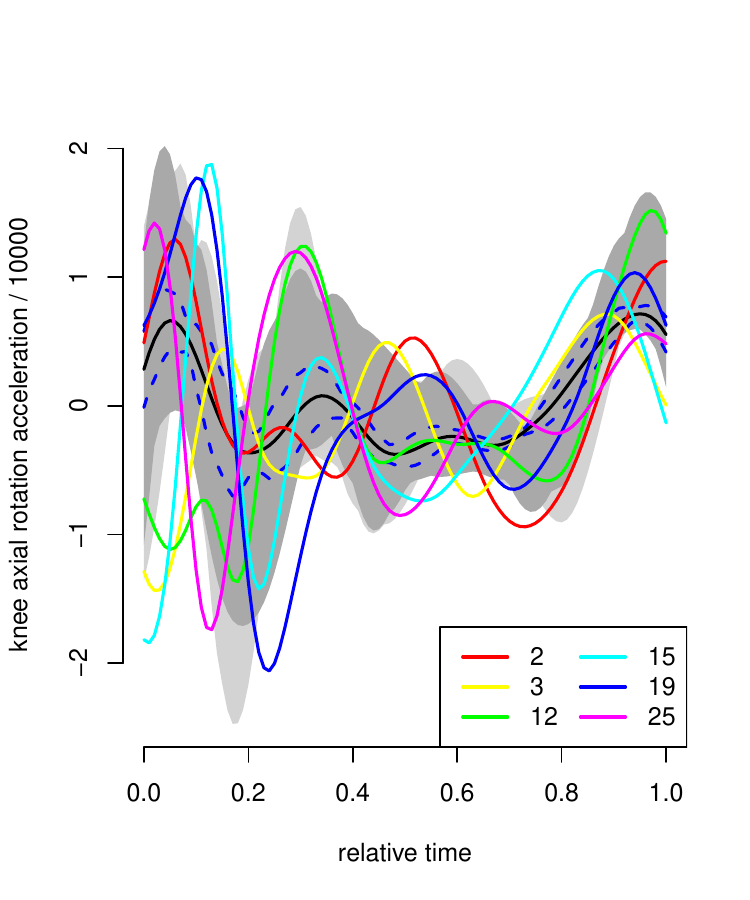}
\caption{Rainbow plot
for the knee axial rotation acceleration profiles from Figure~\ref{fig:examples}~(left) with yellow colors indicating low and red high functional depth; sample median given in black in both plots, with 95\% pointwise confidence intervals (right); functional bagplot with  50\% inner region and 99\% fence region shaded in dark and light gray, respectively, and flagged outlying curves highlighted in color (right). }\label{fig:descr}
\end{center}
\end{figure}

For our running example, Figure~\ref{fig:descr} shows an example rainbow plot and functional bagplot for the knee axial rotation acceleration profiles of Figure~\ref{fig:examples}~(left) obtained using the \texttt{rainbow} R package. Note that the functional depth measure in both plots uses only the first two functional principal components (compare Section~\ref{sec:FPCA}) and thus only captures about 62\% of the variation in the functional data.

\subsection{Smoothing for functional data} \label{sec:smooth}

Functional data is in practice observed on finite grids -- which may be curve-specific -- and commonly with measurement error. A typical model for the observations  thus is
$X_i(t_{il}) = \widetilde{X}_i(t_{il}) + \varepsilon_{il}$,
where $X_i(t_{il})$ constitutes the observable value at $t_{il} \in \mathcal{T}, l=1, \dots, L_i, i=1, \dots, n$, $\widetilde{X}_i(t)$ is the underlying (smooth) function and $\varepsilon_{il}$ are i.i.d.\ noise terms (compare Section~\ref{sec:notation}). As part of an analysis, it is often of interest to reconstruct the underlying ``true'' curves $\widetilde{X}_i(t)$. 

The earliest approaches in functional data analysis usually do this reconstruction as an initial step using presmoothing of the curves \citep{ramsay_functional_2005}. 
For a given curve $i$, we can view 
$X_i(t_{il}) = \widetilde{X}_i(t_{il}) + \varepsilon_{il}$ with i.i.d.\ noise $\varepsilon_{il}$  
as a nonparametric regression problem and use corresponding approaches for scatterplot smoothing. Typical methods used include kernel smoothing, local polynomial smoothing, and spline or penalized spline approaches \citep{ramsay_functional_2005}.

For example, we can expand $\widetilde{X}_i = \widetilde{X}_i(\cdot)$ in a spline basis, i.e., assume that it can be written or approximated (asymptotically) as a linear combination of spline basis functions $\phi_k$, $k=1, \dots, K$,
and write 
\begin{equation}\label{eq:presm}
X_i(t_{il}) = \widetilde{X}_i(t_{il}) + \varepsilon_{il} = \sum_{k=1}^K \phi_k(t_{il}) \theta_{ik} + \varepsilon_{il}.     \end{equation}
Unknown coefficients $\theta_{ik}$ are estimated using a least squares or penalized least squares criterion, penalizing the integrated squared (e.g.\ second) derivative of the function to encourage smoothness or an approximation such as the finite differences of neighboring B-spline coefficients \citep{eilers2021practical}.

After presmoothing, the resulting curves are treated as if they were truly functional observations, i.e., fully observed curves. The advantage of this presmoothing approach is that theory and methods developed for observations that are truly functional \citep[e.g.][]{dauxois1982asymptotic,bosq2000linear} can be directly applied. The clear disadvantage is that the uncertainty in estimating $\widetilde{X}_i(t)$ from noisy and discrete data is ignored in subsequent analysis steps. In particular, for sparse functional data, this source of uncertainty can be substantial and vary across curves depending on the number of available observation points. Also, smoothing can fail or not work well if only a few points for a curve are observed \citep{yao2005functional}.

Thus, more recent approaches have often worked with the observed data and incorporated one of the smoothing methods into the analysis itself, in particular for sparse functional data \citep{yao2005functional}. This has the clear advantage that the true information content of the data is acknowledged both in the analysis and the uncertainty quantification. Also, and importantly, such an approach allows ``borrowing of strength'' across functions, which is particularly useful in the case of sparse functional data.
This approach was spearheaded by \cite{yao2005functional} for the case of functional principal component analysis discussed in Section~\ref{sec:FPCA} below, where smoothing is incorporated in the estimation of the covariance function. Similar ideas were then included in other methods for functional data, such as in regression \citep[e.g.][]{scheipl2015functional,greven2017general,R_refund}. We will introduce these approaches in the corresponding sections. 

A further approach that is similar to presmoothing is to expand the functions in a basis prior to analysis and subsequently work with the multivariate vector of basis coefficients instead of the presmoothed functions. If the basis transformation is loss-less, as can be the case for, e.g., wavelets and functions on regular grids \citep{morris2006wavelet}, this approach can be equivalent to working with the functions (up to potential simplifying assumptions in the subsequent model such as independence between coefficients). For splines \citep{ramsay_functional_2005}, this approach is very similar to presmoothing, and after the denoising usually the resulting uncertainty in curve reconstruction is also ignored.  
Other approaches besides presmoothing that include some sort of denoising as preprocessing of functional data include functional principal component analysis \citep[e.g.][]{goldsmith2011penalized} or factor models \citep{Hoermann2020}.

\begin{figure}[htb]
\begin{center}
\includegraphics[width=70mm]{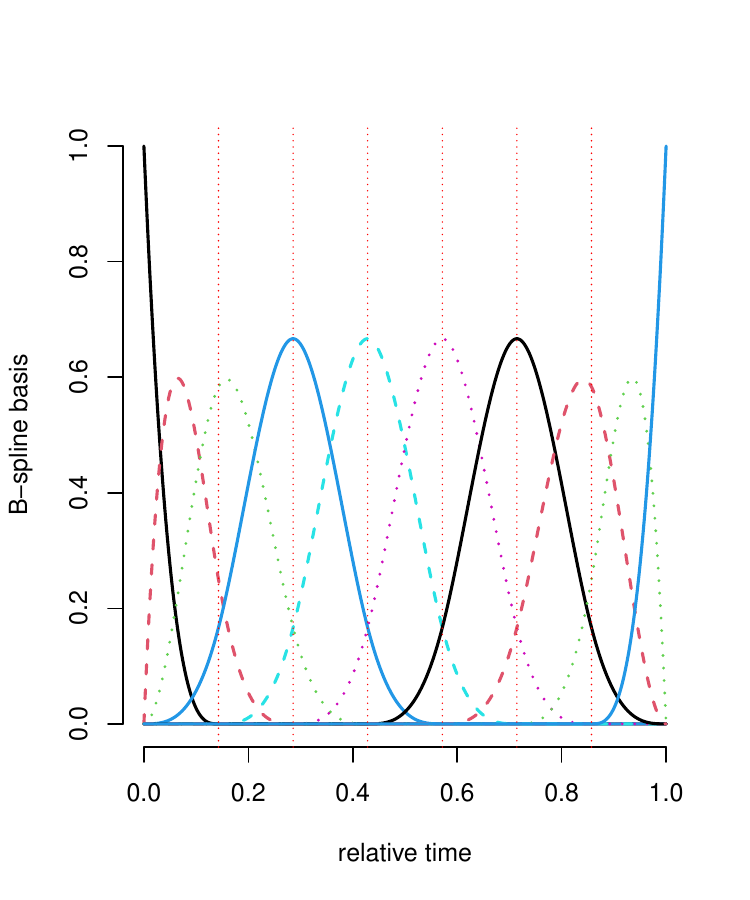}
\includegraphics[width=70mm]{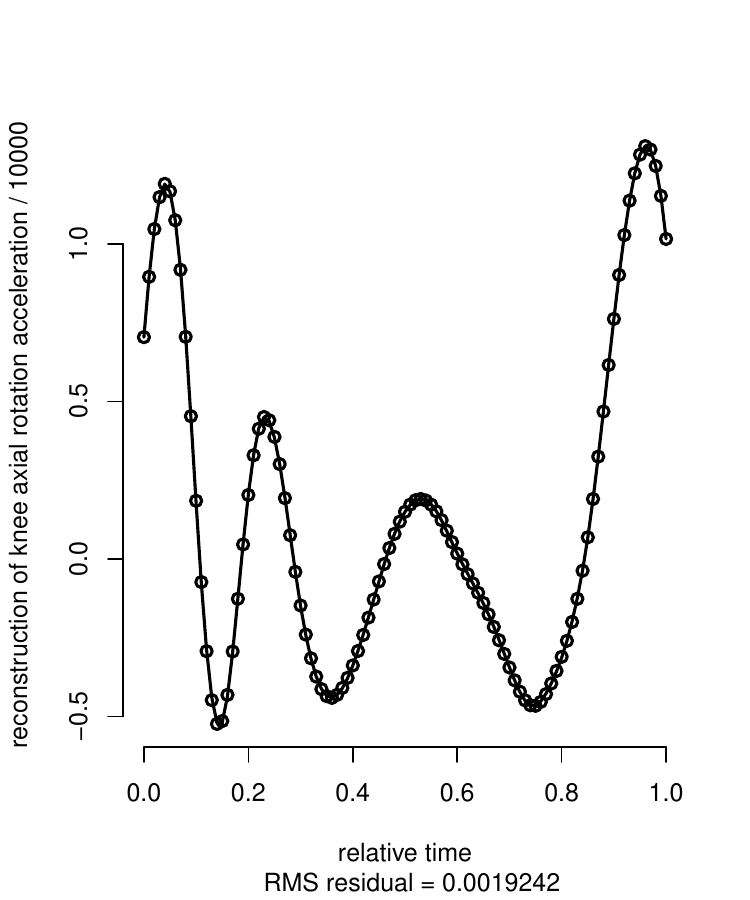}
\caption{B-spline basis with 10 knots (left); 
reconstruction (solid curve) of the first knee axial rotation acceleration profile (observed points as circles) with 30 B-spline basis functions (right). }\label{fig:smooth}
\end{center}
\end{figure}

As data from our running example are already smooth, they can be nearly perfectly reconstructed using a B-spline basis expansion with a sufficient number of knots, e.g., $K=30$. Figure~\ref{fig:smooth} illustrates this for the first knee axial rotation acceleration profile using the \texttt{fda} package. On the left, a basis with only 10 basis functions is shown for better visibility. The reconstruction of the functional observation on the right with 30 basis functions is extremely close to the observed data points.

\subsection{Amplitude and phase variation} \label{sec:phase}

In functional data, we often observe two kinds of variation: \emph{amplitude} and \emph{phase} variation. When displaying functions by plotting $x_i(t)$ versus $t \in \mathcal{T}$, these become visible as variations ``along the y-axis'' and ``along the x-axis'', respectively. Typical examples are functions over time such as, e.g., growth curves \citep{ramsay_functional_2005}, where there is variation both in the height (amplitude) of children as well as in the timing of growth spurts (phase). Further examples are functions over space such as, e.g., diffusion tensor imaging (DTI) measurements along different neuronal tracks in the brain \citep{greven2011longitudinal}, where there is variation both in the fractional anisotropy measurement as a proxy for neuronal health (amplitude), as well as variation in the brain location corresponding to a given $t$ (phase), due to different proportions of anatomies and due to imprecisions in the measurement process. In our running example, phase variability is clearly visible, as maxima and minima in vertical knee acceleration are not aligned in Figures~\ref{fig:examples} and \ref{fig:descr}. This means that there is not only variation in the magnitude of acceleration between observations, but also in the timing of the acceleration pattern.

Ignoring phase variation in an analysis can lead to confounding effects between amplitude and phase and to less interpretable results. Intuitively speaking, in functional data analysis, we usually assume that the same argument $t$ corresponds across functions -- e.g., corresponds to the same event in time or to the same location in space -- and comparisons across functions per $t$ are thus sensible. If there is phase variation, this is no longer the case.
Depending on the context, it is thus often of interest to either (i) align functions before the analysis (of amplitude only) if phase variation is not of interest, which is also known as \emph{registration} or \emph{warping}, or (ii) to separate amplitude and phase variation and analyze them separately or jointly.

Most existing approaches focus on (i), the registration or warping problem. If we want to align the functions $x_i$, e.g., to some overall mean $\mu$, this can be formulated as finding suitable so-called \emph{warping functions} $\gamma_i$ that monotonically map $\mathcal{T}$ onto $\mathcal{T}$, such that $x_i(\gamma_i(t))$ is optimally aligned to $\mu(t)$ for all $t \in \mathcal{T}$. The $\gamma_i$ then capture different ``speeds'' (phase) of the different functions $x_i$ along the interval~$\mathcal{T}$.
Traditionally \citep[e.g.][Ch.~7]{ramsay_functional_2005}, registration has often been performed by matching certain landmarks -- e.g., maxima, minima, zero-crossings, or corresponding values for derivatives -- across functions and linearly interpolating $\gamma_i$ in between. This approach is simple and works reasonably well in the case of clearly defined and equal numbers of such landmarks.
More sophisticated approaches take the whole function into account and often work with minimizing the $\mathcal{L}^2$ distance between one function $x_j$ and a second aligned function $x_i$ over $\gamma$, i.e.,
\begin{equation}\label{L2warping}
 \inf_{\gamma} \| x_i \circ \gamma  - x_j\|^2_{\mathcal{L}^2} =  \inf_{\gamma} \int_{\mathcal{T}} (x_i(\gamma(t)) - x_j(t))^2 dt.   
\end{equation}
 In addition to not being symmetric, i.e., warping $x_i$ to $x_j$ and warping $x_j$ to $x_i$ are not equivalent, this approach carries the so-called \emph{pinching problem} \citep{marron2015}. This means that the distance \eqref{L2warping}  can become small or zero even if $x_i$ and $x_j$ are not warped versions of each other, by using a warping function $\gamma$ that compresses areas of $\mathcal{T}$ where $x_i$ and $x_j$ are dissimilar and expands those where they are close, resulting in spiky warping functions. Regularization or Bayesian approaches 
 \citep[e.g.][]{ramsay_li,ramsay_functional_2005,lu,matuk}
 restrict the amount of warping possible, but can only partially solve this problem. \cite{srivastava} thus proposed to work with the so-called \emph{elastic distance} instead, the Fisher-Rao-metric optimized over warping. This distance can be shown to simplify to the ${\mathcal{L}^2}$ distance in (\ref{L2warping}) for the optimally aligned square-root-velocity transformed curves $Q({x}_i)(t)$ and $Q({x}_j)(t)$, which simplifies computations. Here,
$Q({x}_i)(t) = {\dot{x}_i(t)}/{\sqrt{|\dot{x}_i(t)|}}$  (if $\dot{x}_i(t) \neq 0$, else $Q({x}_i)(t) = 0$) for the first derivative $\dot{x}$ of $x$.
The elastic distance constitutes a proper distance between functions modulo warping (and level due to the derivative) and avoids the pinching problem. It can thus be used to align functions or to obtain e.g.\ a mean estimate after alignment, by taking out all information about phase. Approaches for computations with sparsely observed (multivariate) functional data have also been developed \citep{steyer2022elastic}. 

In some settings, phase is of interest as well. Some more recent approaches thus consider a joint analysis of both phase and amplitude, i.e., point (ii) above, e.g., in the context of functional principal component analysis (cf.\ Section~\ref{sec:FPCA}) \citep{happ2019general} or regression (cf.\ Section~\ref{sec:funreg}) \citep{hadjipantelis2014analysis}.

\begin{figure}[htb]
\begin{center}
\includegraphics[width=48mm]{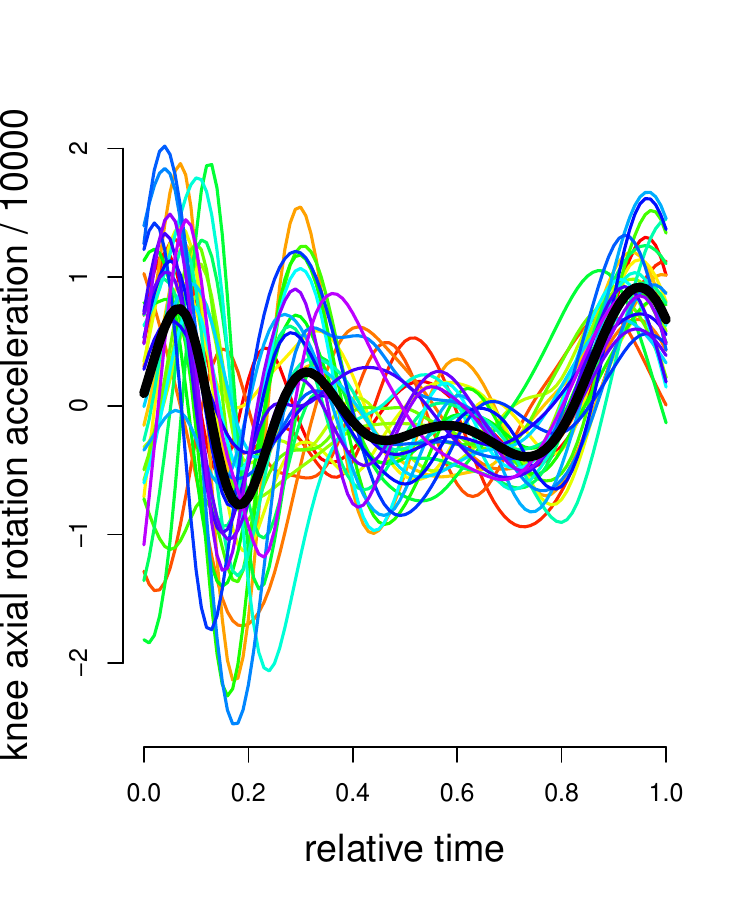}
\includegraphics[width=48mm]{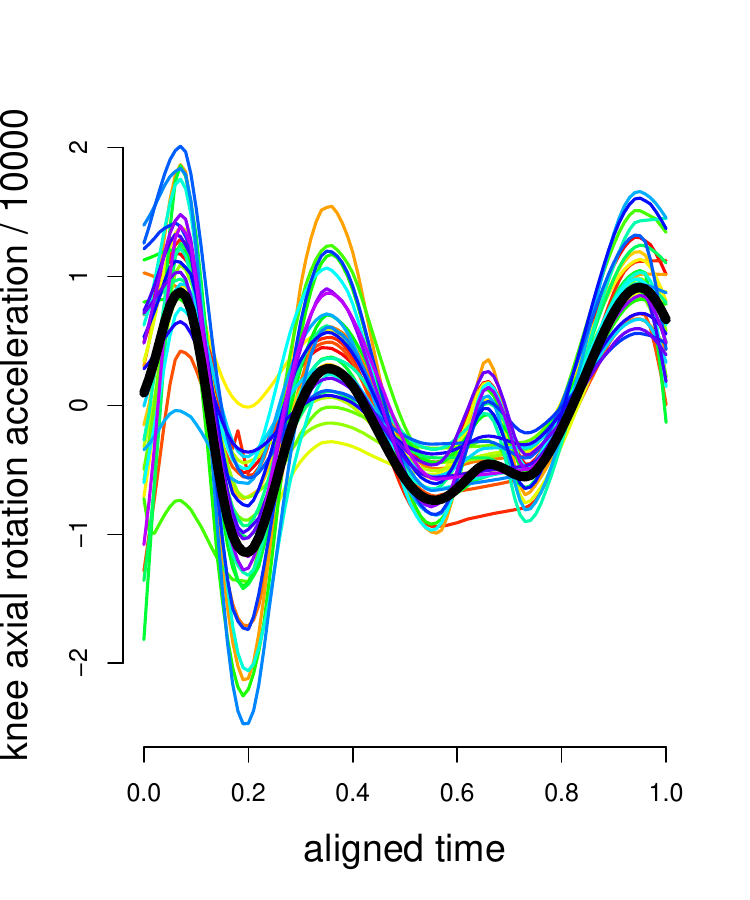}
\includegraphics[width=48mm]{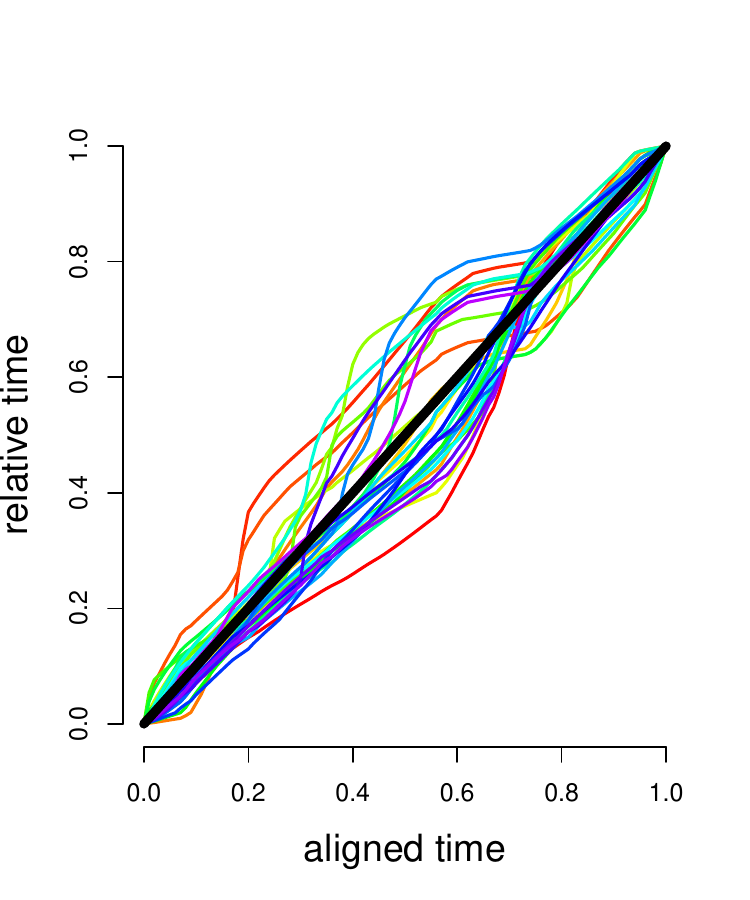}
\caption{Knee axial rotation acceleration profiles from Figure~\ref{fig:examples}~(left) with their mean function in black (left); same functions iteratively aligned to their mean given at the end of the algorithm in black (middle); warping functions warping the original time to the aligned time (right). 
}\label{fig:phase}
\end{center}
\end{figure}

In our running example, there is a clear misalignment of maxima and minima between functions. As the number of maxima and minima can differ between functions, simple warping approaches based on landmarks do not work well here. Figure \ref{fig:phase} shows the knee axial rotation acceleration profiles from Figure~\ref{fig:examples}~(left) before and after elastic alignment using the \texttt{time\_warping} function in R package \texttt{fdasrvf}, together with the warping functions transforming the time for alignment.
Note that warping functions (up to numerics) are centered around the identity line.
Plotted are also the means of the unaligned and aligned functions in black. After the alignment, the mean shows more pronounced peaks and valleys, as it is no longer averaging over functions with their maxima and minima at different time points, and thus better represents the functions in the data. 

\subsection{Functional principal component analysis} \label{sec:FPCA}

Functional principal component (FPC) analysis is an important tool in FDA as it allows dimension reduction of at least in theory infinite-dimensional functional data to manageable finite vectors of FPC scores.
It is based on the Karhunen-Lo\'eve expansion \citep{karhunen1947lineare,loeve}, which allows us to expand a random function, i.e., a stochastic process, $X$ over $\mathcal{T}$ as
\begin{equation}\label{karhunen}
X(t) = \mu(t) + \sum_{m=1}^\infty \xi_m \phi_m(t), \quad t \in \mathcal{T},
\end{equation}
where $\mu$ is the mean function of $X$ and $\phi_m$ are orthonormal eigenfunctions of the covariance, with $\int_{\mathcal{T}} \phi_m(t) \phi_k(t) dt = 1 
$ iff $k=m$ and $=0$ otherwise. In particular, according to Mercer's theorem \citep{mercer1909},
\begin{equation}\label{mercer}
\Cov(X(s), X(t)) =
\sum_{m=1}^\infty \nu_m \phi_m(s)\phi_m(t),
\end{equation}
with decreasing eigenvalues $\nu_1 \geq \nu_2 \geq \dots \geq 0$.
Furthermore, $\xi_m$ are uncorrelated random scores with mean zero and variance $\nu_m$, $m = 1, 2, \dots$, and are independent normal if $X$ is a Gaussian process.

Truncating the sum in \eqref{karhunen} at a finite upper limit $M$ gives the best approximation to $X$ with $M$ basis functions in terms of mean-square prediction error \citep{rice1991estimating}. This allows for a parsimonious representation that approximates a sample  $X_i$, $i=1, \dots, n$, of independent replicates of $X$ with finitely many functions $\phi_m$ common to all observations, and individual scores $\xi_{m,i}$ that capture the function-specific deviations from the overall mean. As the components have decreasing importance as measured by $\nu_m$, the first few components often capture most of the relevant features in the data. Looking at the fraction  $\sum_{m=1}^M \nu_m / \sum_{m=1}^{\infty} \nu_m$ allows a quantitative assessment of the  (integrated) variance explained by the approximation, and a choice of $M$  to explain, e.g., 95\% or 99\% of the overall variance.

Functional principal component analysis (FPCA) estimates $\mu$, the $\xi_{m,i}, \nu_m$ and $\phi_m$ in practice from a sample of curves taken to be realizations $x_i$ of independent copies of \eqref{karhunen}, i.e., $X_i(t) = \mu(t) + \sum_{m=1}^\infty \xi_{m,i} \phi_m(t)$. Usually, a smoothness assumption is incorporated for the estimation of the $\phi_m$, which can be done in several ways. We here briefly describe an approach based on \cite{yao2005functional} that accommodates sparsely observed functions with additional white noise errors $\epsilon_{it}$ and thus works relatively generally. As in this setting presmoothing the observed functions $x_i$ or computing the scores using numerical integration based on $\xi_{m,i} = \int_{\mathcal{T}} (X_i(t) - \mu(t)) \phi_m(t) dt$ does not work well, the idea is to incorporate smoothing into the estimation of the covariance in \eqref{mercer}.  After estimation of $\mu(t)$ by $\hat\mu(t)$, and detrending $\tilde{x}_i(t) = x_i(t) - \hat\mu(t)$, estimation is based on the cross-products of observed points within curves, $\tilde{x}_i(t_{ij})\tilde{x}_i(t_{ik})$, which are roughly unbiased estimates of $\Cov(X(t_{ij}), X(t_{ik}))$. All cross-products are pooled and a smoothing method of choice is used for bivariate smoothing -- local polynomial smoothing in \cite{yao2005functional}, penalized spline smoothing in the \texttt{refund} R package \citep{R_refund}, with further improvements for the particular case of covariance smoothing regarding speed and symmetry proposed e.g.\ in \cite{xiao2013fast,cederbaum2018fast}. If an additional error is assumed, the diagonal cross-products approximate  $\Cov(X(t_{ij}), X(t_{ij}))$ plus the error variance, the diagonal is thus left out for smoothing and the error variance estimated from the difference of the smooth diagonal and the average  (possibly boundary-trimmed) diagonal cross-products. 

Once the smooth covariance is available, the orthogonal decomposition \eqref{mercer} is typically done numerically on a fine grid using the usual matrix eigendecomposition. Eigenvectors are then rescaled to approximate orthonormality of the $\phi_m$ with respect to the functional ${\mathcal{L}^2}$ norm rather than the Euclidean vector norm. Finally, the scores $\xi_{m,i}$ are estimated. In the dense case, this could be based on $\xi_{m,i} = \int_{\mathcal{T}} (X_i(t) - \mu(t)) \phi_m(t) dt$ with the estimated eigenfunctions and numerical integration. In the sparse case, where numerical integration based on only a few grid points does not give good results, it is preferable to work with conditional expectations \citep{yao2005functional} of the $\xi_{m,i}$ given the data. Another way of looking at this is viewing equation \eqref{karhunen} truncated to $M$ eigenfunctions as a linear mixed model with random effects $\xi_{1,i}, \dots, \xi_{M,i}$, which can be assumed to be uncorrelated and thus have diagonal covariance matrix due to the Karhunen-Lo\`eve expansion \citep[e.g.][]{greven2011longitudinal,goldsmith2013corrected,scheipl2015functional}. The eigenfunctions $\phi_m$ take the place of, e.g., a constant function and a linear function for a random intercept random slope model, thus allowing for more modeling flexibility with data-driven basis functions. The conditional expectations then correspond to the usual best linear unbiased prediction of the random effects.

The estimated scores $\boldsymbol{\xi}_i = (\xi_{1,i}, \dots, \xi_{M,i})^\top$ give a multivariate summary vector of the individual functions that provides dimension reduction and enables many methods developed for multivariate data to be transferred to functional data. In addition, the estimated approximation 
$\hat\mu(t) + \sum_{m=1}^M \hat\xi_{m,i} \hat\phi_m(t)$ allows predicting the whole function even from sparse individual data, by pooling of information across functions. \cite{goldsmith2013corrected} provide confidence bands for the individual functions based on this approach that additionally take the uncertainty in the estimated eigendecomposition into account.

\begin{figure}[!htb]
\begin{center}
\includegraphics[width=130mm]{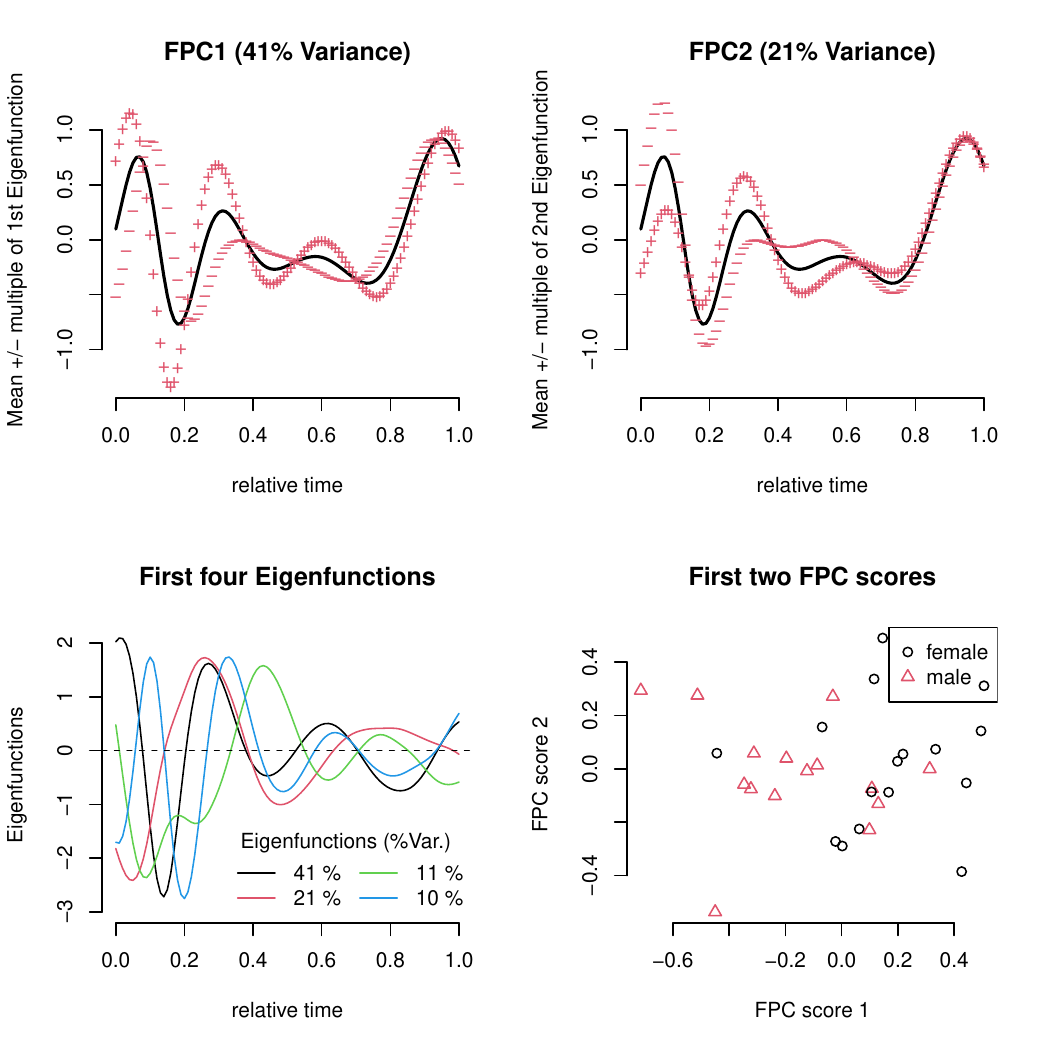}
\caption{First four functional principal components (FPCs, bottom left) 
with the percentage of explained variation in the knee axial rotation acceleration profiles from Figure~\ref{fig:examples}; mean function plus/minus $\sqrt{\hat\nu_1}$ times the first FPC (top left) respectively $\sqrt{\hat\nu_2}$ times the second FPC (top right); scatterplot of second against first FPC scores (bottom right). Note that $\sqrt{\nu_m}$ corresponds to the standard deviation of the $m$th score $\xi_m$, $m=1,2$.
}\label{fig:fpca}
\end{center}
\end{figure}

In our running example, the first four functional principal components explain about 83\% of the total variation of the first knee axial rotation acceleration profiles, as shown in Figure~\ref{fig:fpca}. The importance decreases quickly, with the 5th to 7th FPCs only explaining further 6\%, 4\%, and 2\%, respectively. Positive scores for the first FPC mostly indicate more pronounced movements with higher peaks and lower valleys, and scores for this component tend to be slightly higher for females. The second FPC contrasts more pronounced movements at different times during the whole movement. Please note that FPCs are only unique up to sign, and FPCs and corresponding scores could also be flipped compared to the output of function \texttt{fpca.face} in R package \texttt{refund} in case this allowed for simpler interpretation. We here rescaled the \texttt{fpca.face} output to ensure orthonormality of estimated eigenfunctions with respect to the $\mathcal{L}^2$ instead of the vector inner product, please see the code on Github for details. 
Also note that clearly the eigenfunctions are picking up some of the phase variability in the data that we already saw in Figure \ref{fig:phase}, such that an alternative FPCA would first align the functions and then analyse either the main directions of variation in the aligned data or in the aligned data and warping functions jointly.

Given its centrality as a tool for dimension reduction and to obtain a parsimonious empirical basis, there are several extensions of FPCA to more complex settings. 
Multivariate functional principal component analysis extends FPCA to multivariate functional data. These could be several functions observed over $\mathcal{T}$ such that the setting is that of vector-valued functional data \citep[e.g.][]{chiou2014multivariate}. Or it could be several functions observed potentially over different domains. \cite{happ2018multivariate} develop this setting, with an example of one function being a sparsely observed longitudinal trajectory over time and another being a 2D or 3D image, and provide a link between the univariate and multivariate FPCAs that simplifies estimation.
Covariate-dependent FPCA is, for instance, considered by \citet{ding_functional_2022}.
\cite{di2009multilevel}, \cite{greven2011longitudinal}, and \cite{shou2015structured} develop multilevel FPCA extensions for settings where functions are, e.g., repeatedly observed within observational units such as subjects, and where a FPCA decomposition is desired of both the within-subject and the between-subject functional variation. 
In the case of correlated functional data, we cannot view the $X_i$ as independent copies of \eqref{karhunen}. Extensions of FPCA have thus also been developed, e.g., for functional time series and functions observed repeatedly over time \citep[e.g.][]{chen2012modeling,panaretos2013cramer,park2015longitudinal}, using a double decomposition in both the time dimension and the $t$ dimension.

\subsection{Software}

The Comprehensive R Archive Network (CRAN) \citep{R_core} contains a ``Task View: Functional Data Analysis'' listing most R packages related to functional data analysis. We here name some of the packages implementing methods mentioned in the previous subsections.

For descriptives and outlier detection (Section~\ref{descriptives}), the functional boxplot of \citep{sun2011functional} is implemented in the function \texttt{fbplot} in the R package \texttt{fda}~\citep{fdaR}. The \texttt{rainbow} package~\citep{rainbowR} can be used to obtain functional bagplots, boxplots, and rainbow plots~\citep{hyndman2010rainbow}. \texttt{fdaoutlier}~\citep{fdaoutlierR} provides a collection of functions for functional data outlier detection.

For smoothing of curves (Section~\ref{sec:smooth}), many methods developed for scatterplot smoothing can be used to presmooth functions. The \texttt{fda} package provides functionality to expand observed functions in bases such as, e.g., B-splines. A very flexible R package for smoothing in general is the \texttt{mgcv} package \citep{wood2017gam}. Further packages build smoothing into the analysis, such as, e.g., into the estimation of covariances for FPCA as discussed below.

For warping/alignment of functions and separation of their amplitude and phase (Section~\ref{sec:phase}), elastic options based on the square-root-velocity framework are implemented in R packages \texttt{fdasrvf}~\citep{fdasrvfR} for functions with the same number of sample points and in \texttt{elasdics}~\citep{elasdicsR}  for irregularly or sparsely observed (multivariate) functions. The \texttt{fda} package offers several alternative options based, e.g., on landmarks. 

For FPCA (Section \ref{sec:FPCA}) of univariate functional data, different packages offer options based on covariance smoothing and conditional expectations for the scores. Function \texttt{FPCA} in R package \texttt{fdapace}~\citep{fdapaceR} uses local polynomial smoothing, while several functions in R package \texttt{refund}~\citep{R_refund} use penalized spline smoothing, such as \texttt{fpca.sc} and  \texttt{fpca.face} with the faster FACE method. \texttt{ccb.fpc} provides corrected confidence bands for the predicted functions incorporating uncertainty in estimated FPCs~\citep{goldsmith2013corrected}. R package \texttt{MFPCA}~\citep{MFPCAR} implements FPCA for multivariate functional data~\citep{happ2018multivariate}.

\section{Functional Regression}
\label{sec:funreg}

\subsection{Introduction/Overview}

We call a regression problem `functional regression' if at least one functional variable is found on the left and/or right-hand side of the model equation. This means that functional variables may be the response, covariate(s), or both. To distinguish the different settings, we use the terms `scalar-on-function(s)' regression, `function-on-scalar' regression, and `function-on-function(s)' regression, respectively. In what follows, we will first give an overview of those three settings. Then, we will describe two popular strategies for statistical modeling, inference, and prediction, namely semi-parametric models using basis functions, and non-parametric approaches using kernels. We will focus on the former, semi-parametric models and basis functions here because results are easier to interpret (from our point of view). In addition to the `statistical' approaches described in Section~\ref{sec:funreg}, also methods borrowed from and inspired by the machine learning community can be used, which will be discussed in more detail in Section~\ref{sec:ml}. 

\subsubsection{Scalar-on-Functions Regression (SOFR)} \label{sec:sofr}

`Scalar-on-function(s)' regression corresponds to the case that the response is scalar, and among the covariates, there is at least one functional variable. In the simplest case, there is only a single covariate, which is functional. For instance, Figure~\ref{fig:examples}~(left) showed knee axial rotation acceleration with colors corresponding to the maximum moment. For illustration, we may try to explain/predict the latter using the acceleration profiles. Then the simplest, but probably still the most popular model for scalar-on-function regression is the so-called \emph{functional linear model}
\begin{equation}\label{sof:lin1}
    Y_i = \alpha + \int_\mathcal{T} x_i(t)\beta(t) \, dt + \epsilon_i, \quad i=1,\ldots,n,
\end{equation}
where $\beta$ is the so-called \emph{coefficient function}, $Y_i$ is the (scalar) response and $x_i$ is the functional covariate observed for subject/unit $i$ (such as the curves shown in Figure~\ref{fig:examples}, left). As in the classical linear model with both scalar covariate(s) and response, it is typically assumed that $x_i$ is set by the experimenter, or if the covariate is, in fact, a random variable, we take the `conditional view' in terms of modeling $Y_i$ given $X_i = x_i$. With functional data, however, it is usually the latter. Assumptions for the error term $\epsilon_i$ are typically also the same as in the classical linear model. Thus, $\epsilon_i$ are assumed to be i.i.d.\ (scalar) random variables with mean zero and constant variance $\sigma^2$, and sometimes the additional assumption of normality is made. The latter can, for instance, be relevant when using likelihood-based approaches for estimating unknown model parameters (compare Section~\ref{subsec:semp} below). When interpreting the results after fitting model~(\ref{sof:lin1}) to the data at hand, the coefficient function $\beta$ is particularly important, analogously to the $\beta$-coefficient(s) in the standard linear model, see the examples below.

\begin{figure}[htb]
\begin{center}
\includegraphics[width=70mm]{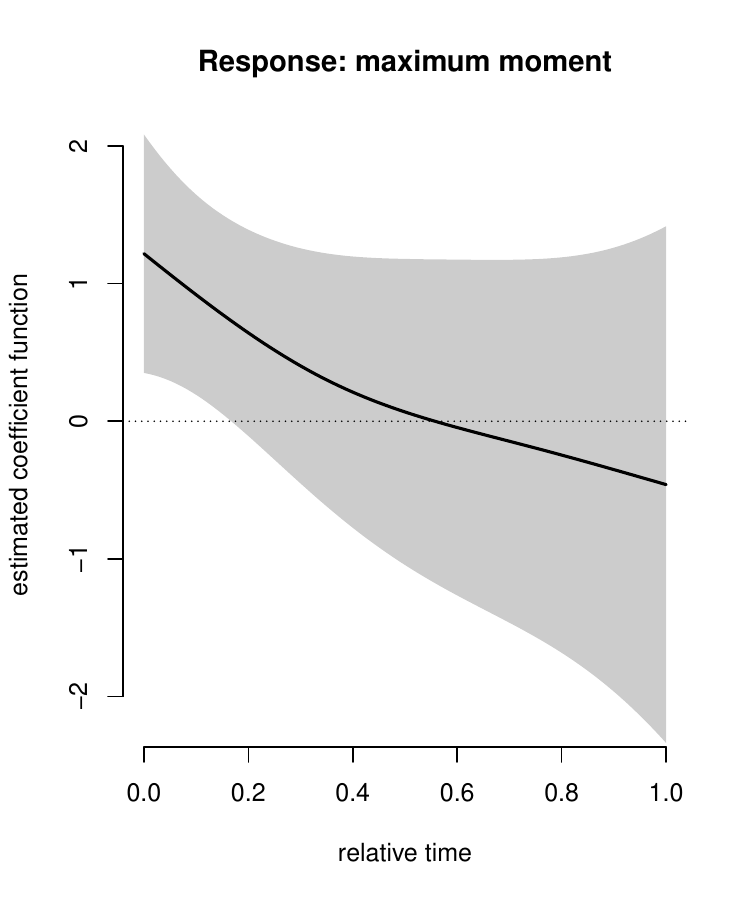}
\includegraphics[width=70mm]{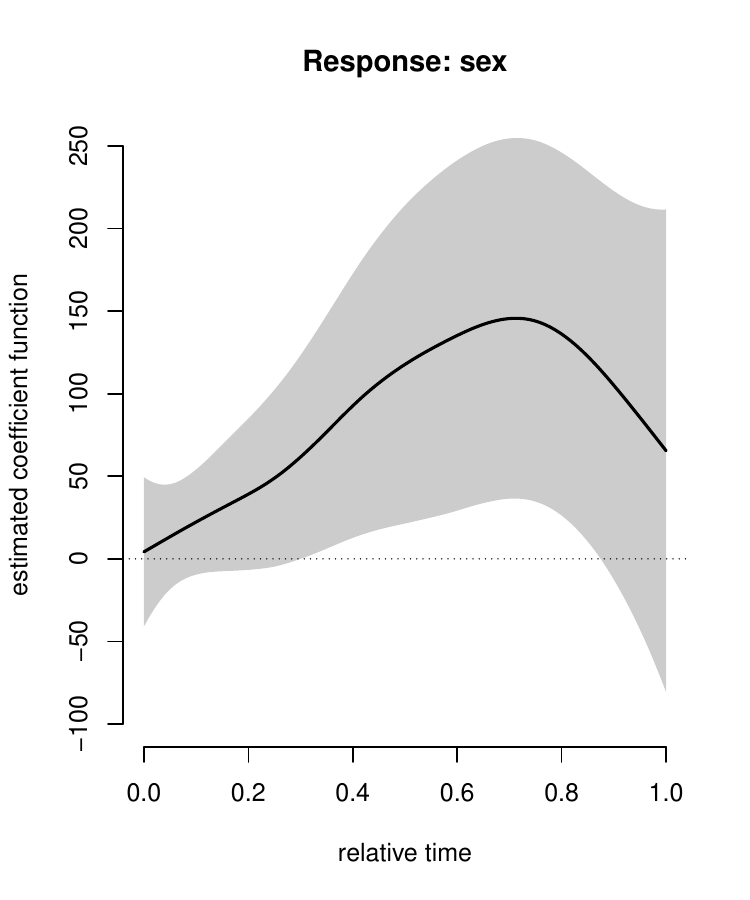}
\caption{The estimated coefficient function in the functional linear model (left) when predicting the maximum moment using the knee axial rotation acceleration profiles from Figure~\ref{fig:examples}~(left) and additional covariate `sex', together with approximate, pointwise 95\% confidence intervals (shaded region). The estimated coefficient function in the functional logit model with response `sex' and functional covariate hip abduction-adduction acceleration from Figure~\ref{fig:examples}~(right) is given in the right panel.}\label{fig:flm}
\end{center}
\end{figure}

As pointed out above, the functional linear model~(\ref{sof:lin1}) is the simplest model for scalar-on-function regression. However, as in the scalar case, it can be seen as the starting point for numerous generalizations and extensions. The most obvious extension is including scalar covariates $Z_1,\ldots,Z_q$ and multiple functional covariates $X_1,\ldots,X_p$ with domains $\mathcal{T}_1,\ldots,\mathcal{T}_p$, respectively, leading to 
\begin{equation}\label{sof:lin2}
    Y_i = \alpha + \sum_{j=1}^p \int_{\mathcal{T}_j} x_{ij}(t)\beta_j(t) \, dt  + \sum_{r=1}^q z_{ir}\gamma_r + \epsilon_i, \quad i=1,\ldots,n,
\end{equation}
with further assumptions being analogous to (\ref{sof:lin1}). In the case of the data shown in Figure~\ref{fig:examples}, for instance, we should at least take the person's sex into account by including a corresponding factor as an additional scalar covariate. When fitting the corresponding model using function \texttt{pfr()} from R package \texttt{refund}~\citep{R_refund}, the coefficient function for the functional covariate `knee axial rotation acceleration' as shown in Figure~\ref{fig:flm}~(left) is obtained, together with approximate, pointwise 95\% confidence intervals (shaded region). For modeling/fitting the coefficient function, we used a penalized, cubic spline with 15 (B-spline) basis functions and a penalty on the second-order differences of neighboring basis coefficients, a so-called \emph{P-spline}~\citep{marx1999psline}. For details on fitting and software, see Sections~\ref{subsec:semp} and \ref{subsec:regr_sw}, respectively. Further statistical inference, such as confidence intervals, is discussed in Section~\ref{sec:inference}. Qualitatively speaking, the interpretation of $\beta(t)$ from Figure~\ref{fig:flm}~(left) is as follows: (positive) values of knee acceleration at the beginning of the cycle have a positive effect on the maximum moment, whereas the association is negative at the end of the cycle, but uncertainty is very high in that area (as can be seen from the large confidence intervals). So, the main takeaway from this example is that people with large (positive) knee axial rotation acceleration at the beginning of the cycle tend to have a larger maximum moment than people with lower acceleration at the beginning of the cycle. This effect is also visible from the colors in Figure~\ref{fig:examples}~(left). In general, however, such findings are not necessarily possible from descriptive plots only, particularly if more than one covariate is present.   

If the error term $\epsilon_i$ in (\ref{sof:lin2}) is clearly non-Gaussian, e.g., because the response is binary, model~(\ref{sof:lin2}) can be generalized to a response variable with simple exponential-family distribution, analogously to the well-known generalized linear model (GLM). With functional covariates, we can define the linear predictor
\begin{equation}\label{sof:glm1}
    \eta_i = \alpha + \sum_{j=1}^p \int_{\mathcal{T}_j} x_{ij}(t)\beta_j(t) \, dt  + \sum_{r=1}^q z_{ir}\gamma_r, \quad i=1,\ldots,n.
\end{equation}
Then, with $\mu_i = \E[Y_i|X_{i1}=x_{i1},\ldots,X_{ip}=x_{ip},Z_{i1}=z_{i1},\ldots,Z_{iq}=z_{iq}]$ being the conditional mean of the response given the covariates, $\eta_i = g(\mu_i)$, and known link function $g(\cdot)$, the \emph{generalized functional linear model}, or \emph{functional generalized linear model} (both terms used in the literature) is obtained. Instead of the link function, the concrete model can be specified through the inverse link, the so-called response function $h = g^{-1}$, in terms of $\mu_i = h(\eta_i)$. Further extensions that relax the assumption of linearity in (\ref{sof:lin1}), (\ref{sof:lin2}), or (\ref{sof:glm1}) will be discussed in more detail in Section~\ref{subsec:semp_nonp} below. A review entirely on scalar-on-function regression is found in \citet{reiss_methods_2017}. For an illustration of the functional GLM (\ref{sof:glm1}), we may try to identify the persons' sex by looking at the hip acceleration profiles as shown in Figure~\ref{fig:examples}~(right). The fitted coefficient function (again using \texttt{pfr()} from \texttt{refund}) in a functional logit model with response `sex' (female: 0, male: 1) is given in the right panel of Figure~\ref{fig:flm}. Again, we used a P-spline approach with 15 basis functions and a second-order penalty on the basis coefficients. In analogy to the usual logit model, the functional version is defined by the link function
\[g(\pi_i) = \log\left\{ \frac{\pi_i}{1-\pi_i} \right\},\]
with $\pi_i = \mu_i = P(Y_i = 1|X_i = x_i)$ being the conditional probability of person $i$ being male given the covariate curve (hip acceleration) $x_i$. Equivalently, the model is specified through the response function 
\[\pi_i = h(\eta_i) = \frac{\exp(\eta_i)}{1 + \exp(\eta_i)}.\]

As seen from Figure~\ref{fig:examples}~(right), males and females can be separated very well by looking at hip acceleration roughly between time points 0.7 and 0.8. Not surprisingly, the corresponding $\beta$-function shown in Figure~\ref{fig:flm}~(right) also has its clear maximum in that area, indicating that subjects with large/positive hip acceleration in that area tend to be men, whereas small/negative values typically correspond to females.

It should be noted, though, that the functional data shown in Figure~\ref{fig:examples} is very smooth and measured on a dense and regular grid. In practice, however, functional data can be noisy and/or sparsely/irregularly sampled. In those cases, functional regression can be combined with (pre-)smoothing as discussed in Section~\ref{sec:fundamentals}, see also Section~\ref{subsec:semp_nonp}, or even joint (Bayesian) modeling of both $Y_i$ and $X_i(t)$ \citep{mclean2013bayesian} to take uncertainty in the smooth underlying functions into account.

\subsubsection{Function-on-Scalar Regression (FOSR)}\label{sec:fosr}

\begin{figure}[htb]
\begin{center}
\includegraphics[width=70mm]{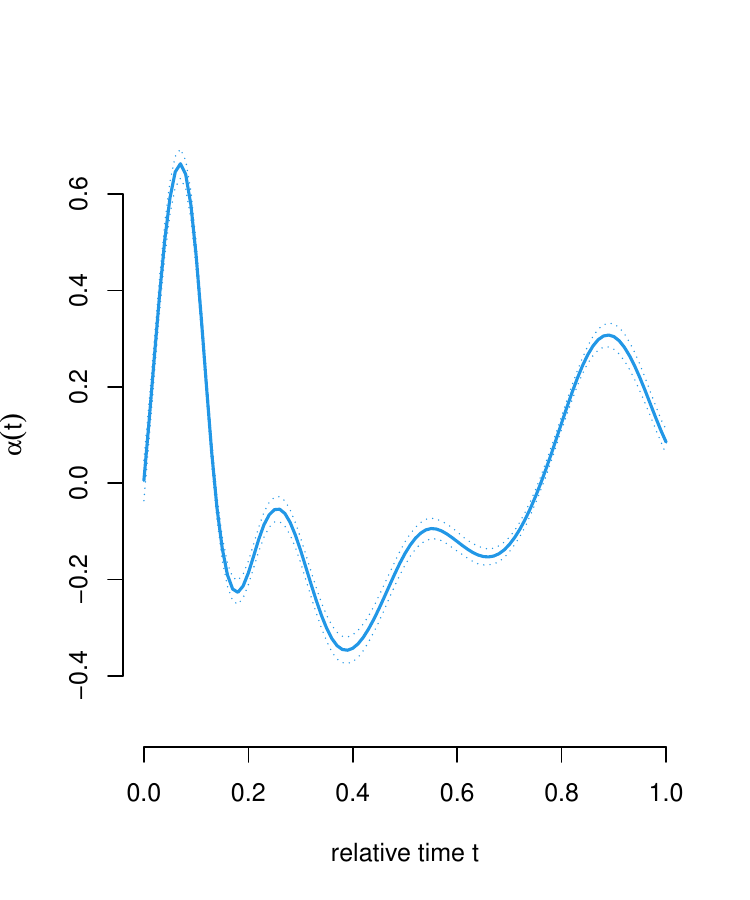}
\includegraphics[width=70mm]{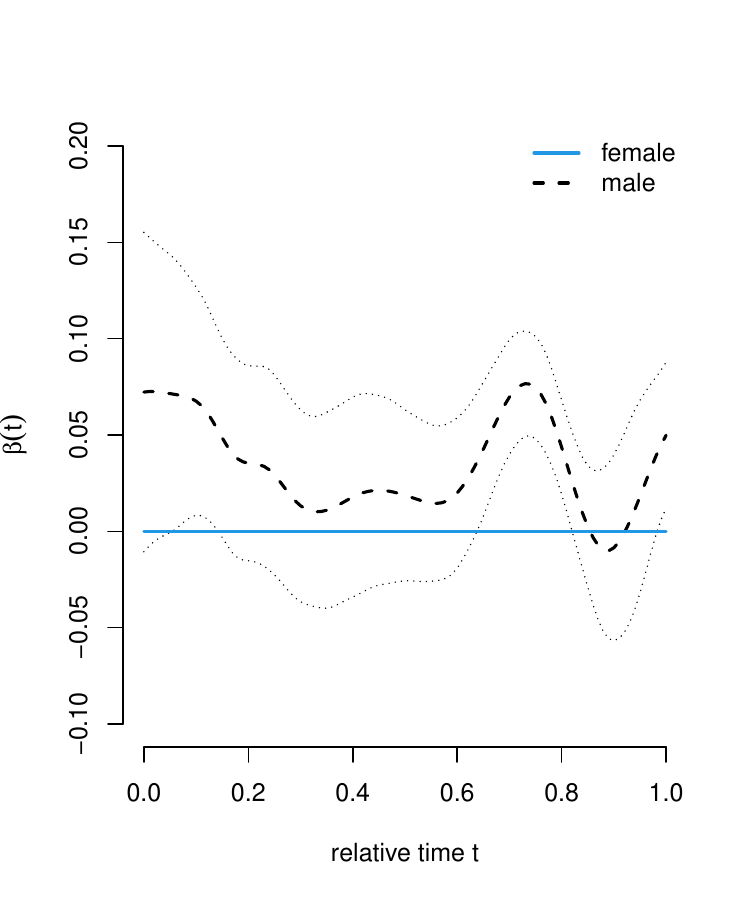}
\caption{Estimated parameter functions when regressing hip abduction-adduction acceleration from Figure~\ref{fig:examples}~(right) on the person's sex using model \eqref{fos:flm}; shown are the functional intercept (left) and the estimated effects (right) with colors corresponding to subjects' sex, and approximate, pointwise 95\% confidence intervals (dotted).}\label{fig:fos}
\end{center}
\end{figure}

If the response $Y(t)$, $t \in \mathcal{T}$ is functional, but all covariates $X_1,\ldots,X_p$ are scalar, the simplest model is the \emph{linear function-on-scalar} model
\begin{equation}\label{fos:flm}
    Y_i(t) = \alpha(t) + \sum_{j=1}^p x_{ij}\beta_j(t) + \epsilon_i(t), \quad i=1,\ldots,n, \; t \in \mathcal{T},
\end{equation}
where $\alpha(t)$ and $\beta_j(t)$, $j=1,\ldots,p$ are coefficient functions and $\epsilon_i(t)$ is the error function drawn from a stochastic process with mean zero and covariance function $\Sigma(s,t)$, $s,t \in \mathcal{T}$; compare, e.g., \citet{chen_variable_2016}. The latter assumption takes into account that measurements taken from the same individual $i$ at different time points, i.e., within-function errors, are typically correlated. This is important for proper statistical inference. \citet{chen_variable_2016}, for instance, apply a ``pre-whitening'' step before estimating unknown regression parameters from \eqref{fos:flm} above. \citet{greven2017general}, by contrast, decompose $\epsilon_i(t)$ into a smooth, subject-specific functional random effect $E_i(t)$ and remaining white noise error $\epsilon_{it}$. For illustration, let us revisit the data from Figure~\ref{fig:examples}~(right), but now treat hip acceleration as the (functional) response and sex as the (scalar) covariate with `female' as the reference category. Then model~\eqref{fos:flm} takes the simple form $Y_i(t) = \alpha(t) + x_{i}\beta(t) + \epsilon_i(t)$,
where $\alpha(t)$ is the functional intercept and $x_i$ is a dummy with $x_i = 1$ if subject $i$ is male, and $x_i = 0$ if $i$ is female. So $\alpha(t)$ (see the corresponding estimate in Figure~\ref{fig:fos}, left) can also be interpreted here as the (functional) mean for females and $\beta(t)$ (estimate in Figure~\ref{fig:fos}, right/black dashed) is the effect of being male. The error process $\epsilon_i(t)$ is decomposed into a smooth, subject-specific functional random effect plus white noise error here~\citep[compare][]{greven2017general}. From Figure~\ref{fig:fos}~(right) it is seen that the first peak in the (mean) acceleration profiles is somewhat higher and wider for males and that males' profiles tend to lie above the profiles of females around $t=0.75$ (also compare Figure~\ref{fig:examples}, right).

In general, besides \eqref{fos:flm}, also more complicated models may be fitted in the semi-parametric framework. For instance, \citet{scheipl2015functional} consider additive, smooth effects in terms of
\begin{equation}\label{fos:fsm}
    Y_i(t) = \alpha(t) + \sum_{j=1}^p \gamma_j(x_{ij},t) + \epsilon_i(t), \quad i=1,\ldots,n, \; t \in \mathcal{T}.
\end{equation}
This means that the effect of covariate $X_j$ is allowed to be non-linear and varying over $t$.

\subsubsection{Function-on-Function Regression (FOFR)}\label{sec:fofr}

If both the response and the covariate(s) are functional and assumed to be observed over the same domain $\mathcal{T}$, a small modification of \eqref{fos:flm} gives the so-called functional linear \emph{concurrent} model as
\begin{equation}\label{fof:fcm}
    Y_i(t) = \alpha(t) + \sum_{j=1}^p x_{ij}(t)\beta_j(t) + \epsilon_i(t), \quad i=1,\ldots,n, \; t \in \mathcal{T}.
\end{equation}
As a modification of \eqref{sof:lin2}, with potentially different domains, we may introduce linear functional effects such that
\begin{equation}\label{fof:flm}
    Y_i(t) = \alpha(t) + \sum_{j=1}^p \int_{\mathcal{S}_j} x_{ij}(s)\beta_j(s,t) \, ds + \epsilon_i(t), \quad i=1,\ldots,n,
\end{equation}
where $t \in \mathcal{T}$, and $\mathcal{S}_j$ denotes the domain of the $j$th functional covariate. Similarly as done in \eqref{fos:fsm} above, \eqref{fof:flm} may be relaxed to \citep{scheipl2015functional}
\begin{equation}\label{fof:fsm}
    Y_i(t) = \alpha(t) + \sum_{j=1}^p \int_{\mathcal{S}_j} F_j(x_{ij}(s),s,t) \, ds + \epsilon_i(t), \quad i=1,\ldots,n,
\end{equation}
although this approach requires the estimation of three-dimensional smooths $F_j$ and is thus less often used in practice.

In particular, if the response and the covariates have the same time domain $\mathcal{T}$, it can also be relevant to allow for more general integration limits than in \eqref{fof:flm}. This allows only (part of) the past of the covariates to affect the current value $Y_i(t)$, but is more general than a concurrent effect as in \eqref{fof:fcm}. Common examples of such \emph{historical} effects replace $\int_{\mathcal{S}_j}$ in \eqref{fof:flm} by $\int_{0}^t$ or $\int_{t-h}^t$ for some lag $h>0$; see \cite{Brockhaus.2017,Ruegamer.2018} for a more detailed discussion of different possible specifications.

\subsection{Semi- and Fully Non-Parametric Approaches}
\label{subsec:semp_nonp}

We will start by giving some further insight into semi-parametric modeling of functional data by use of basis functions, as all models presented so far can be considered in this framework. Fully non-parametric approaches providing even larger flexibility in terms of the association structure between response and covariate(s) will be described in Subsection~\ref{subsec:nonp}.    

\subsubsection{Semi-parametric Models and the Basis Functions Approach}\label{subsec:semp}

The idea here is very similar to presmoothing of functional data as given in~\eqref{eq:presm}. The (unknown) coefficient functions, such as $\beta$ in~\eqref{sof:lin1}, are expanded in basis functions as
\begin{equation}\label{semp:basis1}
    \beta(t) = \sum_{k=1}^K \phi_k(t)b_k,
\end{equation}
where $\phi_k$ could, e.g., be B-spline basis functions as shown in Figure~\ref{fig:smooth} or the eigenbasis obtained via FPCA (compare Section~\ref{sec:FPCA}) of some functional data, typically the covariate functions; and $b_k$ are unknown basis coefficients that need to be estimated from the data. The integrals found in Section~\ref{sec:sofr} and \ref{sec:fofr} are typically calculated/approximated through numerical integration. If considering the functional linear model~\eqref{sof:lin1}, for instance, we have
\begin{equation}\label{semp:approx}
    \int_\mathcal{T} x_i(t)\beta(t) dt \approx \sum_{v=1}^V \Delta(t_v)x_i(t_v)\beta(t_v),
\end{equation}
with a grid $t_1,\ldots,t_V$ of evaluation points in $\mathcal{T}$ and suitable integration weights $\Delta(t_v)$. If evaluation points are equidistant and reasonably dense, the simplest (but most often used) approximation is, up to a proportionality factor, given by $\sum_{v=1}^V x(t_v)\beta(t_v)$. Thus, many methods, such as signal regression as considered by \citet{marx1999psline}, can be seen as functional regression. If choosing the basis functions approach in the (generalized) functional linear model (compare \eqref{sof:lin1}, \eqref{sof:lin2}, or \eqref{sof:glm1}), the functional model turns into a (generalized) linear model with the basis coefficients being the (unknown) regression coefficients that can be estimated using the usual methods for (generalized) linear models. If using basis functions like B-splines, however, typically, a large number of basis functions needs to be used to allow sufficient flexibility concerning the $\beta$-functions that can be fitted. With standard methods like least squares or maximum likelihood, the resulting coefficient functions then tend to over-fit the data while showing some erratic, non-smooth behavior. A popular fix is adding a penalty on the integrated squared (second) derivative(s) of the $\beta$-function(s) or a difference penalty on neighboring basis coefficients to encourage smoothness; compare Section~\ref{sec:smooth} and below. The observed functional data, such as $x_i(t)$ in \eqref{sof:lin1}, might be taken as they are if they are reasonably smooth (compare, e.g., Figure~\ref{fig:flm}), or they may be presmoothed, e.g., by use of pre-chosen basis functions, FPCA, or other techniques (compare Section~\ref{sec:fundamentals}). On the one hand, the use of FPCA has the additional advantage that models \eqref{sof:lin1}, \eqref{sof:lin2} and \eqref{sof:glm1} then transform into a standard (generalized) linear model with FPCA scores as the covariates; compare, e.g., \citet{mueller2005generalized} and \citet{mueller2008additive}. Since often a relatively small number of principal components suffices to capture the main information in the data, the number of unknown regression coefficients in the resulting functional model remains small as well, and further regularization is not necessary. Furthermore, regression on FPCA scores is convenient to generalize. For example, the scores can directly be used in an additive model where the influence of each score is allowed to be non-linear~\citep{mueller2008additive}. On the other hand, however, ``the principal component scores are computed independently from the
response, in an unsupervised fashion, so there is no a priori reason to believe that these scores will correspond to the best dimensions for the regression problem''~\citep{fan2015additive}, and additionally, the discrete parameter $K$ needs to be carefully chosen. Alternatives to additive regression on FPCA scores that induce nonlinear relationships, for instance, include functional index models~\citep{fan2015additive}
\begin{equation}\label{sof:find}
    Y_i = \alpha + \sum_{j=1}^p g_j\left( \int_{\mathcal{T}_j} x_{ij}(t)\beta_j(t) dt\right) + \epsilon_i, \quad i=1,\ldots,n,
\end{equation}
where both $\beta$- and $g$-functions are represented via basis functions and estimated in an iterative manner. Basis coefficients for $\beta$ and $g$ are alternatingly updated while fixing the other set of coefficients at their current values, cycling through those two steps until convergence. Apparently, \eqref{sof:find} includes the functional linear model as a special case if all $g$-functions are set to be the identity. As before, appropriate penalties can or should be included when fitting the basis coefficients of the $\beta$ and/or $g$-functions. 

Analogous basis/penalty approaches can be used for FOSR and FOFR as introduced in Section~\ref{sec:fosr} and \ref{sec:fofr}. In the additive model~\eqref{fos:fsm} or the linear model~\eqref{fof:flm} for function-on-functions regression, however, the unknown functions ($\gamma_j$ and $\beta_j$, respectively) have two arguments. This means that the basis needs to be chosen appropriately, e.g., as a so-called tensor product basis. If considering a single functional covariate in~\eqref{fof:flm}, for instance, equation~\eqref{semp:basis1} then turns into
\begin{equation}\label{semp:basis2}
    \beta(s,t) = \sum_{k=1}^{K_1}\sum_{l=1}^{K_2} \phi_k(s)\phi_l(t) b_{kl},
\end{equation}
and smoothing should typically be done in both $s$- and $t$-direction.

As already pointed out, if using a rich basis such as a large number of B-splines, smoothing is typically carried out by adding a penalty when fitting the basis coefficients (such as those in \eqref{semp:basis1} and \eqref{semp:basis2}), which prevents over-fitting and ensures interpretability of the resulting coefficient functions. In the simplest case, for instance, the functional linear model~\eqref{sof:lin1} with reasonably smooth functional covariate curves that are observed on a dense and regular grid $t_1,\ldots,t_V$, fitting can be done by minimizing the penalized quadratic loss 
\begin{equation}\label{semp:penQ}
    Q(\alpha,\mathbf{b}) = (\mathbf{y} - \boldsymbol{\alpha} - \mathbf{X}\mathbf{\Phi}\mathbf{b})^\top(\mathbf{y} - \boldsymbol{\alpha} - \mathbf{X}\mathbf{\Phi}\mathbf{b}) + \lambda \mathbf{b}^\top\mathbf{\Omega}\mathbf{b}
\end{equation}
as a function of the basis coefficients $\mathbf{b} = (b_1,\ldots,b_K)^\top$ and the intercept $\boldsymbol{\alpha} = (\alpha,\ldots,\alpha)^\top$. Here, $\mathbf{X}$ is a ($n \times V$) data matrix with the entry in row $i$ and column $v$ being $x_i(t_v)$, i.e., $(\mathbf{X})_{iv} = x_i(t_v)$; ($V \times K$) matrix $\mathbf{\Phi}$, with $(\mathbf{\Phi})_{vk} = \phi_k(t_v)$, contains the basis functions $\phi_1,\ldots,\phi_K$ evaluated at $t_1,\ldots,t_V$, $\mathbf{y}$ collects the observed response values, and $\mathbf{\Omega}$ is a penalty matrix specifying the concrete type of quadratic penalty imposed on $\mathbf{b}$. The strength of the penalty is controlled through tuning parameter $\lambda$. Simple matrix algebra shows that, up to integration weights, the $i$th entry of $\mathbf{X}\mathbf{\Phi}\mathbf{b}$ corresponds to \eqref{semp:approx} for equidistant and dense grids. If using a P-spline approach~\citep{marx1999psline} with second-order penalty (as done in Figure~\ref{fig:flm}), for instance, $\mathbf{\Omega}$ takes the form $\mathbf{\Omega} = \mathbf{D}_2^\top\mathbf{D}_2$, where $\mathbf{D}_2$ produces second-order differences in coefficients as $\mathbf{D}_2\mathbf{b} = (b_3 - 2b_2 + b_1,\ldots,b_K - 2b_{K-1} + b_{K-2})^\top$. In the case of a generalized model, such as \eqref{sof:glm1}, the quadratic loss $(\mathbf{y} - \boldsymbol{\alpha} - \mathbf{X}\mathbf{\Phi}\mathbf{b})^\top(\mathbf{y}  - \boldsymbol{\alpha} - \mathbf{X}\mathbf{\Phi}\mathbf{b})$ in \eqref{semp:penQ} is replaced by the negative log-likelihood. If more than one (functional) covariate is to be considered (compare~\eqref{sof:lin2}), the design matrix $\mathbf{X}$ as well as the matrix of basis functions $\mathbf{\Phi}$ and vector of basis coefficients $\mathbf{b}$ needs to be extended accordingly, and further penalty terms are added. Since suitable values for the associated penalty parameter(s) are typically unknown, those need to be chosen in a data-driven way. Besides generic approaches such as cross-validation or information criteria, a popular and smart way in the case of \emph{quadratic} penalties (as they are typically used with functional data), is offered through the \emph{mixed model} perspective. The latter means that quadratic smoothing penalties are equivalent to interpreting basis coefficients as random effects with normal (prior) distribution and implies a one-to-one correspondence of variance components and smoothing parameters. As a consequence, the latter can be estimated via maximum likelihood or restricted maximum likelihood; compare \citet{wood2011fast}, \citet{wood2016smooth}, \citet{wood2017gam}, and \citet{scheipl2015functional, scheipl2016generalized}. In addition to the estimation of smoothing parameters, the mixed model perspective also offers tools for further statistical inference, such as testing and confidence intervals (compare Section~\ref{sec:inference}).       

An important point when choosing the final model is variable selection, particularly if the number of potential covariates is large. As before, we need to distinguish between scalar-on-function, function-on-scalar, and function-on-function regression. Most contributions in the literature, however, are on the first two cases, with a very popular approach being sparsity-inducing penalties; see, e.g., \citet{matsui_variable_2011}, \citet{gertheiss_variable_2013}, \citet{feng_variable_2022} for variable selection in the (generalized) linear model for scalar-on-functions regression, \citet{fan2015additive} for the functional index model; and \citet{lian_shrinkage_2013}, \citet{chen_variable_2016}, \citet{barber_function--scalar_2017}, \citet{fan_high-dimensional_2017}, \citet{parodi_simultaneous_2018} for the function-on-scalar case. Alternatives include functional lars~\citep{cheng_nonlinear_2020}, Bayesian methods~\citep{goldsmith_smooth_2014, kowal_bayesian_2020}, and boosting as discussed in Section~\ref{sec:ml}. Also, more classic approaches such as forward/backward selection based on statistical testing as given in Section~\ref{sec:inference} could be used. A recent review on variable selection in functional regression models is provided by \citet{aneiros_variable_2022}. Finally, 
variable selection with functional data is sometimes interpreted as selecting the most predictive design (grid) points~\citep{ferraty_most-predictive_2010, blanquero_variable_2019} or intervals/regions in the functions' domain~\citep{james_functional_2009, tutz_feature_2010, zhou_functional_2013, centofanti_smooth_2020}; which typically requires alternative methods to the basis functions and smoothing approach as described above. For further information, we refer to the literature cited above. The following subsection provides a short introduction to fully non-parametric functional regression.

\subsubsection{Non-parametric Approaches}\label{subsec:nonp}

Let us consider the regression problem with continuous $Y_i$ and a single functional covariate $X_i$ in its most general form, where
\begin{equation}\label{sof:nonp}
Y_i = f(X_i) + \epsilon_i,
\end{equation}
$f$ being an unknown regression function, and $\epsilon_i$ some mean zero noise variable, potentially with some further assumptions such as being i.i.d.\ across subjects $i = 1,\ldots,n$. Compared to the semi-parametric approaches above, in particular, the functional linear model~\eqref{sof:lin1}, assumptions with respect to both the form of $f$ and the distribution of $\epsilon_i$ are much milder.

For functional as for multivariate covariates, for a new observation with known covariate value $x$ and unknown $Y$, a kernel-based, non-parametric prediction $\hat{Y} = \hat{f}(x)$ is given by 
\begin{equation}\label{sof:npregr1}
\hat{f}(x) = \frac{\sum_{i=1}^{n} Y_i K(d(X_i,x)/h_n)}{\sum_{i=1}^{n} K(d(X_i,x)/h_n)},
\end{equation}
with some kernel $K$, bandwidth $h_n\searrow 0$ (for $n \rightarrow \infty$) and distance measure $d$ appropriate for the type of covariate considered, as further discussed 
below. From the mean zero error term, it follows that $\hat{f}(x)$ is an estimate of the conditional expectation $\E[Y|X=x]$. For $\hat{f}$ as defined above to be consistent, a necessary condition, informally speaking, is that the probability of a training observation to be in the close neighborhood of $x$ in terms of $d$ is strictly positive. For details on this so-called \emph{small ball probability}, see, e.g., \cite{FerratyVieu}. In general, each density function may be used as kernel $K$, but typically we restrict ourselves to symmetric kernels with the maximum at zero, such as the (standard) normal density -- the so-called \emph{Gaussian} kernel. For details on kernel functions, see, e.g., \cite{GasMueMam1985}.      

When dealing with functional covariates, in particular, the choice of $d$ is crucial. A popular choice for functional $X_{i},x \in \mathcal{L}^2$, for instance, is the $\mathcal{L}^2$ norm as introduced in \ref{sec:notation},
\begin{equation}\label{sof:dmetric}
d(X_{i},x) = \| X_{i} - x \|_{\mathcal{L}^2}.  
\end{equation}
However, restricting $d$ to be a metric such as~\eqref{sof:dmetric} is sometimes too restrictive in the functional case. So-called \emph{semi-metrics} may also be considered such as
\begin{equation}\label{sof:dsemi}
d(X_i,x) =  \| \dot{X}_{i} - \dot{x} \|_{\mathcal{L}^2}, 
\end{equation}
where $\dot{X}_i,\dot{x}$ are (first) derivatives, see \cite{FerratyVieu} for a deeper insight into this topic. An important difference between metrics and semi-metrics is that in the latter case $d(X_i,x) = 0$ does not necessarily imply $X_i = x$. With~\eqref{sof:dsemi}, for example, $d(X_i,x) = 0$  is also obtained if $x(t) = X_i(t) + c$, for some vertical shift constant $c \neq 0$. 
In general, the choice of which (semi-)metric to take depends on the shape of the data and the goal of the statistical analysis. For instance, if dimension reduction for functional observations is of interest, one possibility is FPCA as discussed in Section~\ref{sec:FPCA}, in which case options for $d$ include distances on the extracted scores. In general, results can look very different, depending on the chosen measure of proximity. In Chapter~3 of \cite{FerratyVieu} examples to illustrate this effect are given. Also, further suggestions for semi-metrics and a survey on which semi-metric may be appropriate for which situation can be found there. For example, semi-metric~\eqref{sof:dsemi}, which is based on the derivatives, is often well suited for smooth data, whereas for rough data a different approach should be considered.

\begin{figure}[htb]
\begin{center}
\includegraphics[width=70mm]{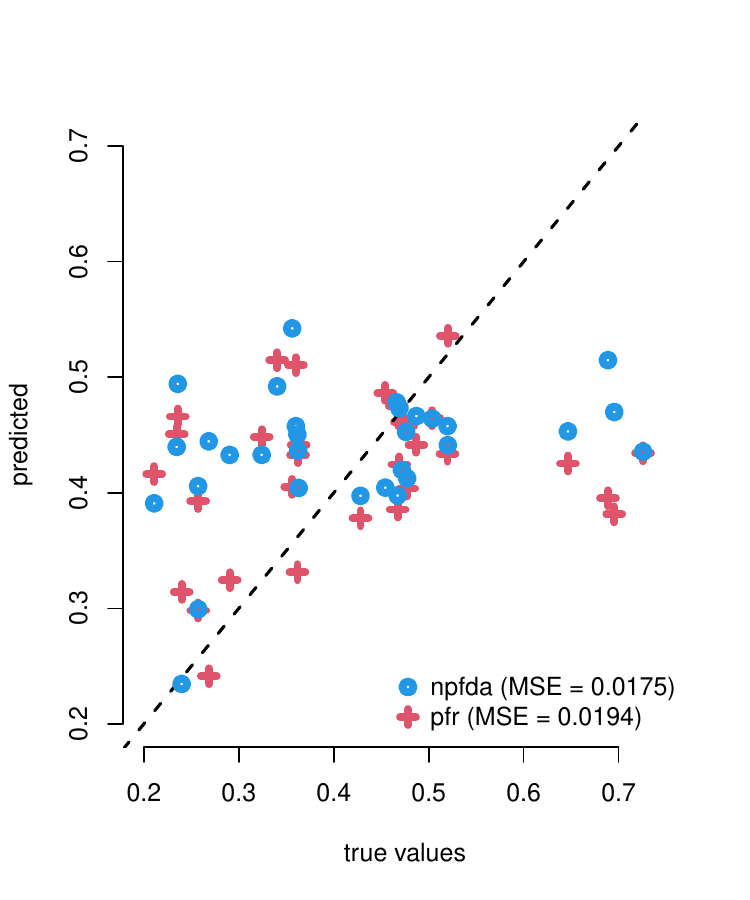}
\includegraphics[width=70mm]{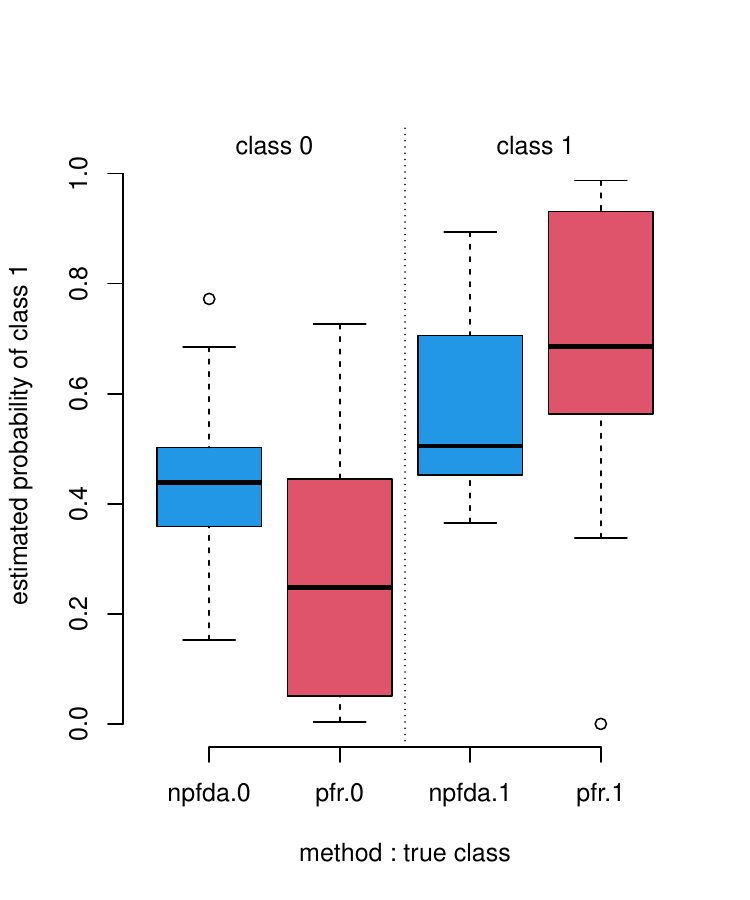}
\caption{Predicted values vs.~true values (left) and estimated probabilities of class 1 (i.e., `male') vs.~true classes (right; $0 =$ `female', $1 =$ `male') when using nonparametric functional data analysis (`npfda') or a (generalized) functional linear model fitted by \texttt{pfr()} on the data from the left and right of Figure~\ref{fig:examples}, respectively.}\label{fig:npfda}
\end{center}
\end{figure}

For an illustration of nonparametric scalar-on-function regression~\eqref{sof:npregr1}, consider the knee acceleration data from Figure~\ref{fig:examples}~(left). We use nonparametric regression/prediction~\eqref{sof:npregr1} with a semi-metric in terms of an (unweighted) Euclidean distance on the scores from an FPCA with 5 components. The (global) bandwidth is chosen via (inner) cross-validation. Figure~\ref{fig:npfda}~(blue circles, left) shows the predicted values obtained via leaving-one-out (outer) cross-validation vs.~the true values. For comparison, we also give the results for a functional linear model (fitted by \texttt{pfr()}, and applying leaving-one-out cross-validation as well). A perfect prediction would correspond to values on the bisecting (dashed black) line. We see that this is a rather difficult regression problem, with nonparametric functional data analysis (`npfda') performing slightly better in terms of the mean squared error (MSE), which indicates at least some non-linearity in the association between the response and the covariate (although the larger MSE of `pfr' is mainly caused by the two extreme observations with true values around 0.7 that are fitted much better by `npfda'). A potential issue with nonparametric functional regression as introduced so far, however, is that it is restricted to cases with only one (functional) covariate. An approach how to extend~\eqref{sof:npregr1} to settings with additional, functional, or scalar covariates (such as `sex'), is presented in \citet{selk2023nonparametric} and \citet{selk2023asymp}. 

If using the nonparametric approach~\eqref{sof:npregr1} on the data from Figure~\ref{fig:examples}~(right) with a binary, 0/1-coded response (female: 0, male: 1), the conditional expectation/regression function $f(x)$ corresponds to the (conditional) probability of class 1 given input $x$. The corresponding estimates obtained by npfda and pfr are shown in Figure~\ref{fig:npfda}~(right). As we can see, the functional logit model from Section~\ref{sec:sofr} performs much better here than the nonparametric approach. This can be explained as follows. As noted earlier (compare Figure~\ref{fig:examples}, right), the two classes can be separated very well by focusing on hip acceleration between time points $0.7$ and $0.8$. The coefficient function fitted by \texttt{pfr()} is able to exploit this (compare Figure~\ref{fig:flm}, right), whereas the distance/semi-metric $d$ used here (FPCA) is not. The nonparametric approach might perform better, for instance, by introducing an appropriate weight function in~\eqref{sof:dmetric}. Examples of such functions, including an algorithm for estimating a weight function from the data at hand, are given by \citet{CheReiTar2014}. An advantage of the nonparametric approach~\eqref{sof:npregr1} is that it can be easily used for regression with a multi-class response as well. As this is closely related to the classification of functional data, we refer to Section~\ref{subsec:class} for more details. Furthermore, the kernel-based approach presented in Section~\ref{subsec:nonp} could be combined with the more structured models as presented in \ref{sec:sofr}. For instance, \citet{jeon2021additive} proposed a very general approach for additive regression with nonstandard response and covariates, including functional data.

\subsection{Software}\label{subsec:regr_sw}

Some functions for functional regression analysis, where either the dependent variable or one or more independent variables are functional, are available in R package \texttt{fda}~\citep{ramsay_functional_2005, ramsay_funRMatlab_2009}; in particular, for the functional linear model for scalar-on-function regression~\eqref{sof:lin2} and the concurrent model~\eqref{fof:fcm}. A broader class of models, including the models presented in \ref{sec:sofr} to \ref{sec:fofr}, can be fit by use of \texttt{refund}~\citep{R_refund}. The functions implemented there typically use \texttt{mgcv} methodology~\citep{wood2011fast, wood2016smooth, wood2017gam}, and the latter R package can also be used directly to fit at least some of those models, but the use of \texttt{refund} may often be easier for the inexperienced user. R functions for nonparametric functional data analysis as presented in~\ref{subsec:nonp} and thoroughly discussed in the monograph by \cite{FerratyVieu} are found at the accompanying website \url{http://www.math.univ-toulouse.fr/staph/npfda/}.

\section{Statistical Inference with Functional Data}
\label{sec:inference}

\subsection{Functional ANOVA}

\begin{figure}[!htb]
\hspace{-3mm}\includegraphics[width=154mm]{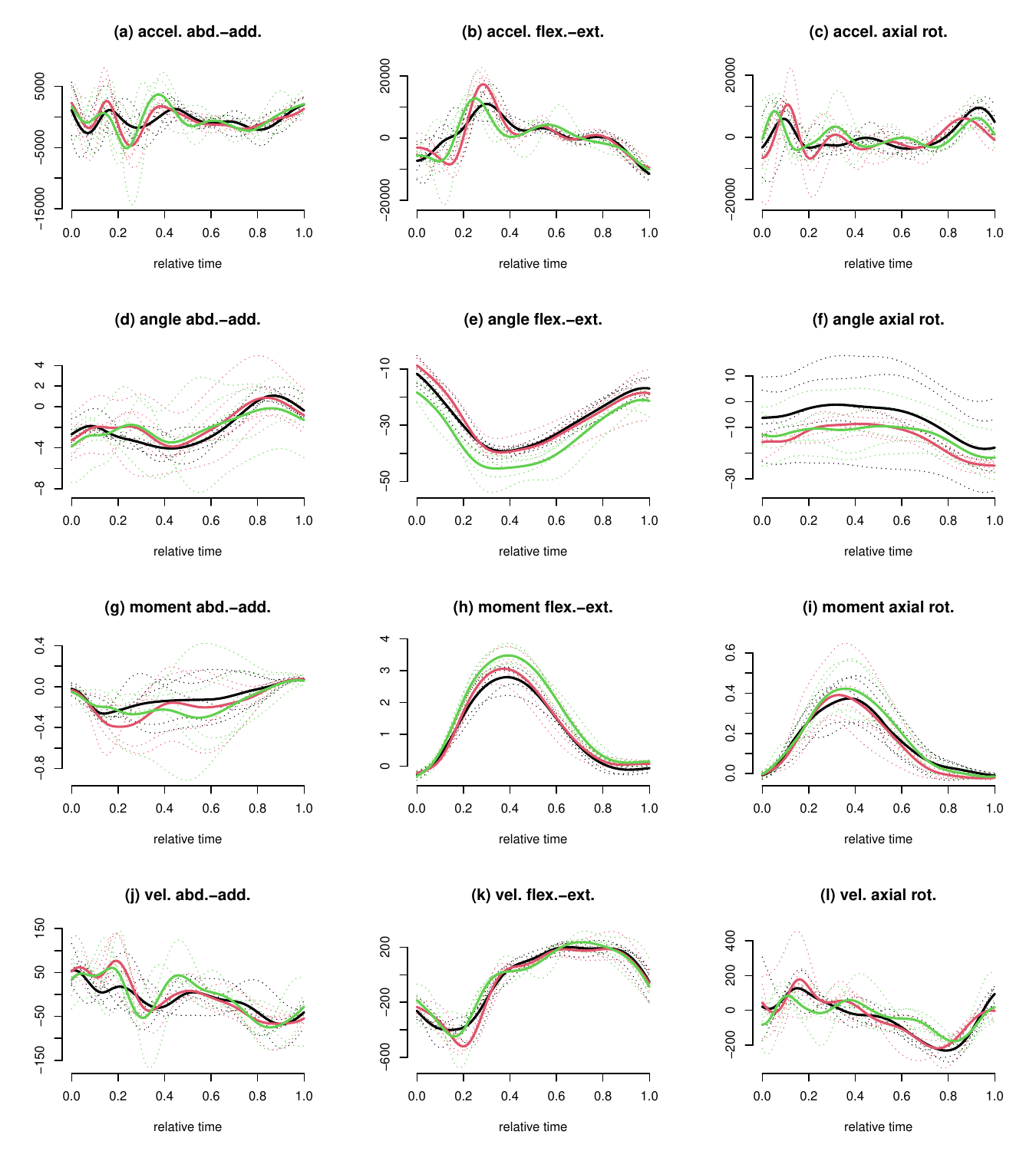}
\caption{Example knee measurements (dotted lines), compare Table~\ref{tab:vars}, for three experimental conditions: `slowbw' (black)  `slowten' (red), `slowtwe' (green); the solid curves give group-wise means.}\label{fig:fanova1}
\end{figure}

The setting of one-way univariate functional ANOVA is similar to the well-known scalar case. We typically have a grouping variable, a so-called factor, defining groups $1,\ldots,g$, and functional data $X_{ij}(t)$, $t \in \mathcal{T}$, coming from these $g$ groups, $j=1,\ldots,g$, $i=1,\ldots,n_j$, with $n_j$ denoting the sample size in group $j$. The overall sample size is $n = n_1 + \ldots + n_g$. Given group membership, functional variables are assumed to be (conditionally) independent. Furthermore, it is typically assumed that functions are square-integrable on $\mathcal{T}$, i.e., elements of $\mathcal{L}^2(\mathcal{T})$ (compare Section~\ref{sec:notation}). The hypothesis to test is
\begin{equation}\label{fanova:H0a}
    H_0: \mu_1(t) = \ldots = \mu_g(t), \quad t \in \mathcal{T},
\end{equation}
with $\mu_j(t)$ denoting the mean function in group $j$. Alternatively, we can use a linear function-on-scalar model with one categorical covariate in terms of
\begin{equation}\label{fanova:m1}
    X_{ij}(t) = \mu(t) + \alpha_j(t) + \epsilon_{ij}(t),
\end{equation}
where $\mu(t)$ is the grand mean (function), $\alpha_j(t) = \mu_j(t) - \mu(t)$ the deviation from the grand mean in group $j$, and $\epsilon_{ij}(t)$ a mean zero error/noise process. Furthermore, it is typically assumed that all functions $X_{ij}(t)$, and thus $\epsilon_{ij}(t)$, share the same covariance function $\Sigma(s,t)$, $s,t \in \mathcal{T}$. This is the functional version of homoscedasticity as commonly assumed in the scalar case (see below for literature on testing the equality of covariance operators between groups). For identifiability, a constraint such as $\sum_j \alpha_j(t) = 0$ for all $t$ is needed. Then, hypothesis~(\ref{fanova:H0a}) can be rewritten as
\begin{equation}\label{fanova:H0b}
    H_0: \alpha_1(t) = \ldots = \alpha_g(t) = 0, \quad t \in \mathcal{T}.
\end{equation}
An example is given in Figure~\ref{fig:fanova1}. Here, the data (dotted lines), a subset of our running example data, comes from $g=3$ groups defined by different experimental conditions: run at 3 m/s with no backpack (`slowbw'), with 10\% body mass backpack (`slowten'), with 20\% body mass backpack (`slowtwe'). The empirical, group-wise mean functions are given as solid lines. For simplicity, we restrict ourselves to a small, illustrative data set comprising 15 males, with five persons in each group.\\

In what follows, we will give a short overview of some procedures proposed for testing~(\ref{fanova:H0a}) and (\ref{fanova:H0b}), respectively. Many of these tests are based on the pointwise between-groups sum of squares
\begin{equation}
    \text{SSR}(t) = \sum_{j=1}^g n_j(\hat{\mu}_j(t) - \hat{\mu}(t))^2
\end{equation}
and the pointwise within-groups/error sum of squares
\begin{equation}
    \text{SSE}(t) = \sum_{j=1}^g \sum_{i=1}^{n_j} (X_{ij}(t) - \hat{\mu}_j(t))^2,
\end{equation}
compare \citet{ramsay_functional_2005} and \citet{gorecki_fdanova_2019}. One way of constructing a test statistic then is the ratio of the integrated between- and within-groups sum of squares, that is
\begin{equation}\label{fanova:Rat}
    \frac{\frac{1}{g-1} \int_\mathcal{T} \text{SSR}(t) dt}{\frac{1}{n-g} \int_\mathcal{T} \text{SSE}(t) dt},
\end{equation}
and the tests proposed in the literature vary concerning the method used for representing the functional data and estimating the unknown parameters, and the way p-values are calculated. For instance, under some assumptions such as Gaussian error processes, \citet{shen_2004} derived the test statistics' null distribution and also provided an approximation based on the Karhunen-Loéve expansion of the error process. \citet{zhang_2011} proposed an improved, biased-reduced approximation, and \citet{gorecki_fdanova_2015} used a finite-dimensional basis function representation of the observable functional data $X_{ij}(t)$ as the starting point.

As an alternative to \eqref{fanova:Rat}, we may also calculate the pointwise F-statistic,
\begin{equation}\label{fanova:F}
    F(t) = \frac{\frac{1}{g-1} \text{SSR}(t)}{\frac{1}{n-g} \text{SSE}(t)},
\end{equation}
and integrate/globalize this quantity in terms of $\int_\mathcal{T} F(t) dt$~\citep{zhang_2014}, or use only its maximum, i.e., $\sup_{t \in \mathcal{T}} F(t)$~\citep{zhang_2019}. 

A somewhat different approach is presented in \citet{cuevas_2004}, who used the sum of pairwise ${\mathcal{L}^2}$-distances $\sum_{j<l} \int_\mathcal{T} (\hat{\mu}_j(t) - \hat{\mu}_l(t))^2 dt$ as the test statistic and a parametric bootstrap to approximate the null distribution. \citet{cuesta_2010}, by contrast, considered testing through random projects of the functional data observed. Specifically, a trajectory $v$ is sampled on a fine grid, e.g., through a sequence of partial sums of independent normal variables (i.e., a Gaussian random walk). Then, the functional data is projected as $P_{ij} = \int_\mathcal{T} X_{ij}(t)v(t) dt$, and a test is performed for mean differences of the projected data between groups. This procedure is repeated $k$ times, i.e., for $k$ different $v$-trajectories, and resulting p-values are corrected using the FDR approach of \citet{benjamini_2001}. For testing mean differences in the (scalar) projections $P_{ij}$, there are several options, for instance, the approach by \citet{brunner_1997}

\begin{figure}[htb]
\centering
\includegraphics[width=153mm]{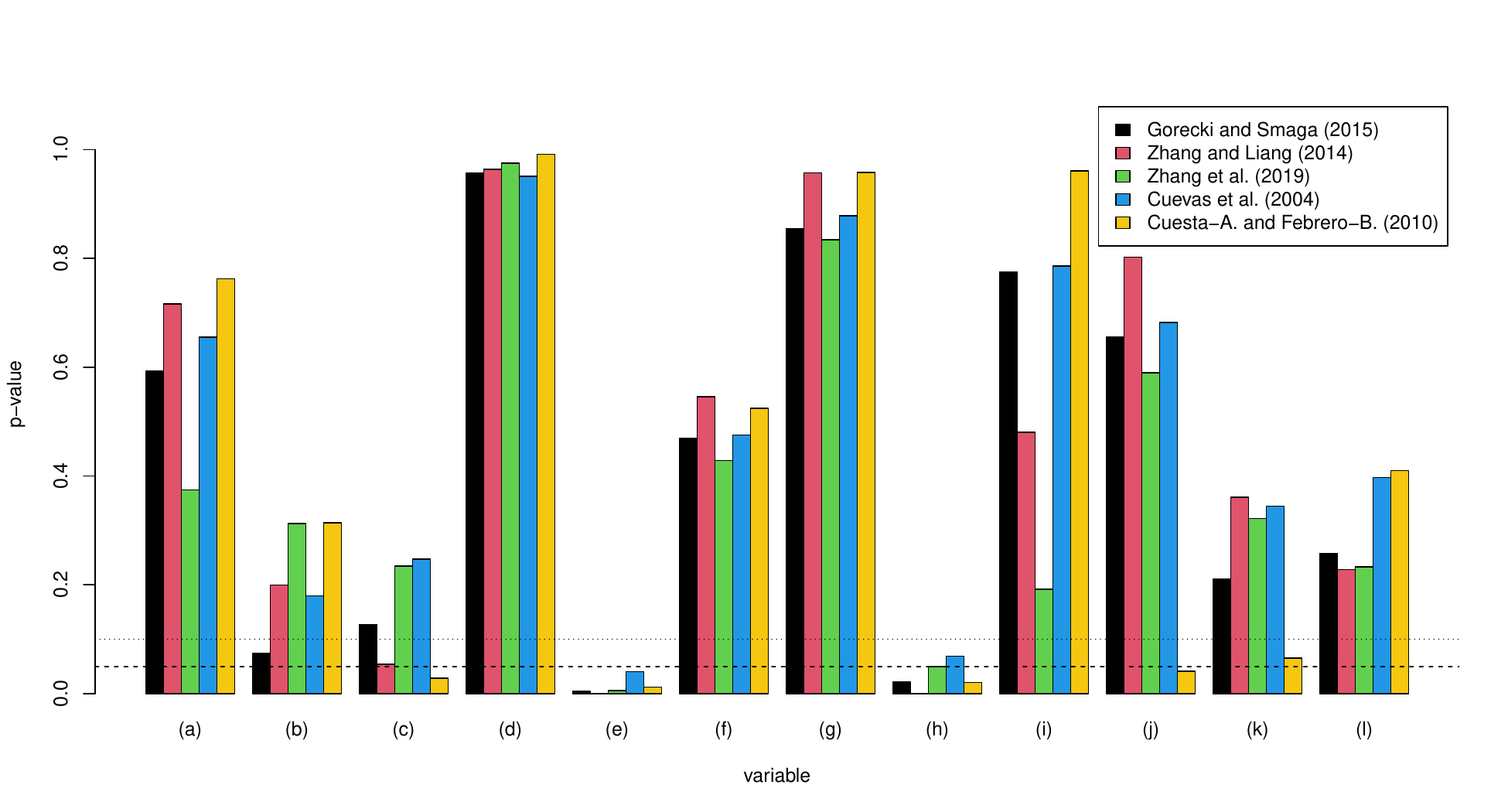}
\caption{P-values for five different tests for functional ANOVA on the data shown in Figure~\ref{fig:fanova1}; significance on the 5\% or 10\% level is indicated by the dashed and dotted line, respectively.}\label{fig:fanova2}
\end{figure}

For illustration, Figure~\ref{fig:fanova2} shows the p-values for five different tests for functional ANOVA on the data shown in Figure~\ref{fig:fanova1}. Computations were done using R package \texttt{fdanova}~\citep{gorecki_fdanova_2019}, which provides about a dozen tests for testing~\eqref{fanova:H0a}. Here we considered a permutation test employing the ratio of the integrated sums of squares~\eqref{fanova:Rat} and basis representation of the functional data~\citep{gorecki_fdanova_2015}, globalizing the pointwise F-statistic~\eqref{fanova:F}~\citep{zhang_2014}, using its maximum only~\citep{zhang_2019}, ${\mathcal{L}^2}$-distances between groupwise means~\citep{cuevas_2004}, and random projections~\citep{cuesta_2010}. Figure~\ref{fig:fanova2} shows large agreement in terms of significant differences between groups (`yes' or `no', according to the dashed/dotted lines in Figure~\ref{fig:fanova2}). In particular, significant differences are detected between the three experimental conditions for (e) `knee flexion-extension angle' and (h) `knee flexion-extension moment'. The different tests, however, are obviously not equivalent. For example, the projection-based approach also identifies significant differences (at the 5\% level) in the mean functions of (c) `knee axial rotation acceleration' and (j) `knee abduction-adduction velocity', whereas the other tests do not. In light of these findings, and keeping in mind that a large number of tests is available for testing \eqref{fanova:H0a}--even many more than considered in Figure~\ref{fig:fanova2}, see~\citet{gorecki_fdanova_2019}--it should be pointed out that running all these tests on the data, and reporting significant differences if any test has a p-value below $\alpha$, is not admissible, since it may substantially increase the type-I error rate. For illustration, we used all data available~\citep{liew.2021} for knee abduction-adduction angle, drew a random sample of 150 curves, and used these curves to form, also at random, five groups of equal size (30). Thus, the null hypothesis~\eqref{fanova:H0a} is true by design because all curves are sampled from the same population. We then repeated this procedure a thousand times and stored the p-values given by the five tests from Figure~\ref{fig:fanova2}. As a result, each test, if considered separately, indeed controls the type-I error rate (or may even be conservative), but the naive approach of considering all five tests together and reporting significance if any p-value is below $\alpha$ increased rejection rates to 8.7\% and 17.1\% well above the nominal level of 5\% and 10\%, respectively. An obvious solution to this problem is to decide on the test to use \emph{before} analyzing the data. But even if we do, there might still be issues, particularly if we choose to use the approach based on random projections~\citep{cuesta_2010}. Due to the randomness involved, it can be very difficult to replicate the results obtained, even on the exact same data. For instance, if running the test on (c) `knee axial rotation acceleration' (Figure~\ref{fig:fanova1}) a hundred times, we obtained p-values smaller than 5\% (as seen in Figure~\ref{fig:fanova2}, yellow) only in about 10\% of all runs. In addition, this provides another easy way of ``cheating'' to researchers who are merely ``hunting statistical significance''~\citep{szucs_2016, simmons_2011}: they may just run the test a few times (e.g., with a sequence of some popular seeds for the random number generator such as `1234', `2345', etc.\ for ``reproducibility'') until the desired result appears. For instance, if allowing for five trials in a setting with all curves actually sampled from the same population described above, we observed type-I error rates of around 10\% for nominal $\alpha$ of 5\%. Consequently, although it is tempting, researchers should refrain from such behavior.\\     

An important special case of \eqref{fanova:H0a} is the two-sample problem, where just two groups, such as case/control or treatment/control, are to be compared; and many testing procedures have been proposed for such situations; see, e.g., \citet{ramsay_functional_2005}, \citet{ZhaLiaXia:2010}, \citet{HorKokRee:2013} and \citet{GhiIevPag:2017}. Besides mean functions, the distributions of functional data from different groups may also be compared in a broader sense~\citep[e.g.,][]{BenHaeKne:2009, pomann_two-sample_2016, pini_itp_2016, KrzSma:2021, wynne_kernel_2022}. Furthermore, there are proposals targeting specifically the covariance functions~\citep[e.g.,][]{KraPan:2012, FreEtal:2013}. An entire textbook on inference for functional data is, for instance, provided by \citet{HorKok2012}.

The comparison of mean functions as in testing~\eqref{fanova:H0a} can be generalized in (at least) two ways. First, we may consider more than one factor in multi-way (functional) ANOVA; see, e.g., \citet{cuesta_2010}. Second, the functional variables to compare between groups may be multivariate, which yields a functional version of MANOVA~\citep[e.g.,][]{gorecki_fdmanova_2017}.

\subsection{Testing and Confidence Intervals in Functional Regression}

In the framework of function-on-scalar regression, various testing procedures can be found in the literature. For instance, after rewriting the linear model~\eqref{fof:flm} as $Y_i(t) = \mathbf{x}_{i}^\top\boldsymbol{\beta}(t) + \epsilon_i(t)$, ${\mathcal{L}^2}$-norm based~\citep{zhang_statistical_2007} and F-tests~\citep{shen_2004, zhang_2011} have been proposed for testing the general linear hypothesis $H_0: \mathbf{C}\boldsymbol{\beta}(t) = \mathbf{c}(t)$, where $\mathbf{C}$ is a given matrix of appropriate dimension, and $\mathbf{c}(t)$ is a corresponding vector of given functions. Functional ANOVA as in \eqref{fanova:H0a} can be seen as a special case of this more general testing problem.

The opposite setting is scalar-on-function regression, where a scalar response is regressed on a functional covariate. In its most general form \eqref{sof:nonp}, only a few assumptions are made, while particularly in the functional linear model \eqref{sof:lin1}, assumptions are much stronger. The most interesting hypothesis to test in this setting is typically whether there is an effect of the covariate on the response at all. That means, in \eqref{sof:nonp}, the null hypothesis is that the regression function $f$ is constant. In the linear model \eqref{sof:lin1} this translates into the coefficient function being zero. Also, if comparing \eqref{sof:lin1} and \eqref{sof:nonp}, an interesting hypothesis to test could be whether the linear model \eqref{sof:lin1} holds. For both problems, a few different testing procedures have been proposed. For instance, \citet{swihart_restricted_2014} and \citet{mclean_restricted_2015} used the mixed model perspective of functional regression and proposed restricted likelihood ratio tests for checking the significance of a functional covariate and/or linearity of the effect. \citet{KonStaMai:2016} extended four tests common in classical regression (Wald, score, likelihood ratio, and F-tests) to the functional linear model by use of FPCA, intending to test the null hypothesis that there is no association between a scalar response and a functional covariate. A Wald-type test is also discussed in \citet{su_hypothesis_2017}, and \citet{yi_f-type_2022} present an improved F-test. \citet{garcia-portugues_goodness--fit_2014} proposed a goodness-of-fit test based on random projections for the functional linear model with scalar response that can also be used for testing the special case of no association between the functional covariate and the scalar response. A thorough comparison of at least some of those tests is found in \citet{tekbudak_comparison_2019}.

\bigskip

In addition to the fitted coefficient functions in a (generalized) functional linear model, Figure~\ref{fig:flm} also provides pointwise 95\% confidence intervals as shaded regions. Those were estimated using the \texttt{pfr()} function from \texttt{refund}, which uses \texttt{mgcv} methodology (compare Section~\ref{subsec:regr_sw}), and are based on the Bayesian view of the smoothing process. For simplicity, let us consider the basis coefficients $\mathbf{b}$ from \eqref{semp:penQ} where only a single functional covariate was present. Given the data and smoothing parameter, the posterior distribution of $\mathbf{b}$ is multivariate normal with some covariance matrix $\boldsymbol{\Gamma}_\mathbf{b}$ (for details on the form of $\boldsymbol{\Gamma}_\mathbf{b}$, see, e.g., \cite{wood2017gam}). Then, with $\hat{\boldsymbol{\beta}} = (\hat{\beta}(t_1),\ldots,\hat{\beta}(t_V))^\top = \boldsymbol{\Phi}\hat{\mathbf{b}}$ denoting the vector that contains the estimated coefficient function at the evaluation points, an approximate pointwise ($1-\alpha$)100\% credible interval for $\beta(t_v)$ is given by $\hat{\beta}(t_v) \pm z_{1-\alpha/2}\sqrt{\zeta_v}$, where $\boldsymbol{\zeta} = (\zeta_1,\ldots,\zeta_V)$ is the diagonal of $\boldsymbol{\Phi}\boldsymbol{\Gamma}_\mathbf{b}\boldsymbol{\Phi}^\top$ and $z_{1-\alpha/2}$ is the ($1-\alpha/2$) quantile of the standard normal distribution. These intervals also have ``surprisingly good frequentist coverage properties'' (averaged over $t$) \citep{wood2017gam}. Since $z_{0.975} = 1.96 \approx 2$, $\hat{\beta}(t_v) \pm 2\sqrt{\nu_v}$ gives an approximate pointwise 95\% confidence interval. When calculating confidence intervals in function-on-scalar or function-on-function regression models, it is important to make sure that within-function correlation is accounted for (compare Section~\ref{sec:fosr}). This may, for instance, be done in a conditional model through functional random effects~\citep{scheipl2015functional} or a marginal model through robust standard errors~\citep{chen_marginal_2013, gertheiss_marginal_2015}.

While pointwise confidence intervals (only) work `pointwise' in terms of coverage properties, so-called simultaneous confidence bands can be interpreted globally. That means the provided bands cover the \emph{entire} (true) function with some prespecified probability such as 95\%. Most methods proposed for simultaneous confidence bands for functional data use resampling techniques, particularly different versions of the bootstrap \citep[e.g.,][]{degras_scb_2011, chang_scb_2017}. Some approaches also use dimension reduction through functional principal component analysis and the Karhunen-Loève expansion; e.g., \citet{goldsmith2013corrected} and \citet{choi_reimherr_2018}. In a very recent paper, \cite{liebl_reimherr_2023} present a broad framework based on random process theory. Among other things, the confidence bands proposed also provide some `local' control of the false positive rate over subsets of the bands' domain.

\subsection{Software}

A collection of various tests for functional ANOVA is found in R package \texttt{fdANOVA}~\citep{gorecki_fdanova_2019}, and all p-values shown in Figure~\ref{fig:fanova2} were calculated through the \texttt{fanova.tests()} function provided there. The R package \texttt{fdatest}~\citep{fdatestR} implements the interval testing procedure by \citet{pini_itp_2016}. Pointwise confidence intervals for terms in SOFR, FOSR, and FOFR models are, for instance, provided by \texttt{refund} functions \texttt{pfr()} and \texttt{pffr()}, respectively. Both are wrappers for \texttt{mgcv}'s \texttt{gam()} and its siblings (compare Section~\ref{subsec:regr_sw}).

\section{Classification and Clustering} \label{sec:beyond}

The purpose of regression, as presented in Section~\ref{sec:funreg}, was to explain and/or predict a scalar or functional response variable using one or more scalar/functional explanatory variables. 
Further approaches with origin in the machine learning community and a clear focus on prediction will be discussed in Section~\ref{sec:ml}. In Section~\ref{sec:beyond}, we will present two further topics beyond regression for functional data from a statistical perspective, namely classification, and clustering. We will start with classification, which shows close links to Section~\ref{sec:funreg}.

\subsection{Classification of Functional Data}
\label{subsec:class}
We assume that functional data comes from distinct classes $1,\ldots,G$, and training data with known class labels is available for determining a classification rule such that new (functional) data with unknown group membership can be assigned to its underlying class with low probability of an error. According to Bayes' rule, each unit should be classified based on posterior class probabilities, that is, the conditional probabilities of each class given the (functional) data. In practice, however, those conditional probabilities are unknown and need to be estimated from the training data. In the simplest case, functional data only come from two classes. Then, we are in a situation like Figure~\ref{fig:flm}~(right), where we can simply use the functional logit model to estimate conditional class probabilities and assign each unit to the class of the highest (estimated) probability.

If $G > 2$ the semiparametric approach, in particular the functional logit model, can be generalized in terms of a functional multinomial model~\citep{matsui_multi_2014}, increasing, however, the number of parameter functions that need to be estimated. The nonparametric, kernel-based approach as presented in Section~\ref{subsec:nonp} on the other hand, can be applied directly. Specifically, following the idea of nonparametric regression~\eqref{sof:npregr1}, we can estimate the conditional probability $P_g(x):=P(Y=g|x)$ of class $g$ given a (functional) covariate $x$ with unknown class label $Y$ by
\begin{equation}\label{sof:npregr2}
\hat P_g(x)=\frac{\sum_{i=1}^nI_g(Y_i)K(d(X_{i},x)/h_n)}{\sum_{i=1}^nK(d(X_{i},x)/h_n)}.
\end{equation}
Here $I_g(Y_i)$ denotes the indicator function, which equals one if $Y_i = g$ and zero otherwise. Besides the semi- and nonparametric approaches described so far, there are many further methods that could be used for the classification of functional data. After dimension reduction via FPCA, for instance, the obtained scores may be used as input to any statistical or machine learning algorithm designed for the classification of multivariate data, e.g., a random forest or a neural net. Also, functional data may be used directly in some situations, e.g., as input to a convolutional neural network. Please refer to Section~\ref{sec:ml} for details. A recent review on functional data classification is provided by \cite{wang2023funclass}.

\subsection{Clustering}
\label{subsec:clust}

In~\ref{subsec:class} above, functional data came from distinct classes $1,\ldots,G$, which were known for the training data and unknown for the test data. The purpose was to classify the test data using a classification rule that was learned from the training data. Since there are known class labels for the training data, this type of learning problem is also called \emph{supervised} learning. If those classes are not given but are to be built using the (functional) data only, we call this process `clustering', which is a so-called \emph{unsupervised} learning problem. Further supervised and unsupervised learning techniques for functional data from a machine learning perspective are discussed in Section~\ref{sec:ml}. The basic idea of clustering is that data are grouped, such that those groups, the `clusters', are homogeneous in some sense. A survey on functional data clustering is, for instance, provided by~\citet{Jacques.2014}. Here, we will only give a short, and not necessarily exhaustive overview.

In general, most clustering methods (not only functional clustering) can roughly be classified as either `data/observation-based' or `model-based'. In the latter case, it is assumed that the data observed are generated by some underlying stochastic mechanism. Then, observations can, e.g., be clustered according to conditional class probabilities. Data-based methods, by contrast, make no distributional assumptions but start directly from the data. Common strategies involve hierarchical clustering and k-means (or variants thereof, such as k-median or k-medoids). Hierarchical clustering algorithms are typically distance-based and either successively group similar objects into clusters and fuse clusters (`agglomerative' hierarchical clustering), or subsequently split larger into smaller clusters (`divisive' hierarchical clustering). The concrete form of the algorithm results from the chosen distance measures for objects and clusters. In both cases, a `hierarchy' of partitions is created from which clusters might be chosen. For details on hierarchical clustering, see, e.g., \citet{murtagh_hierarchical_2014}. As an alternative, we may try to optimize some criterion across all potential partitions into $k$ clusters. The latter is done with k-means, etc.; and a couple of different algorithms have been proposed for doing so (\citealp{MacQueen_1967}; \citealp{hartigan_algorithm_1979}; just to name two of the most prominent references). For an overview of clustering algorithms (including categorizations apart from data vs.\ model-based) see, e.g., \cite{ezugwu_automatic_2021} and references therein.

With functional data, we suggest classifying clustering algorithms according to the \emph{two} dimensions `data/observation-based' vs.~`model-based', \emph{and} `raw data methods' vs.~`dimension reduction', because any combination may be possible. In the case of raw data methods, curves are clustered on the basis of their evaluation points. Within the `dimension reduction' category, we may further distinguish between `filtering methods' and `adaptive methods'~\citep{Jacques.2014}. Filtering methods first approximate the functions by using a set of basis functions and then use the basis coefficients for clustering. Adaptive methods, by contrast, carry out dimension reduction and clustering simultaneously, which may involve that also the type and number of basis functions may depend on the cluster. Model-based functional clustering typically assumes a mixture model that is based on some kind of basis expansion. For instance, \citet{james_clustering_2003} use a spline basis and assume that basis coefficients follow cluster-specific Gaussian distributions. \citet{bouveyron_model-based_2011} and \citet{jacques_funclust_2013} allow for cluster-specific eigendecompositions (FPCA). Similarly to usual, model-based clustering for multivariate data, various specifications for the underlying Gaussian distributions are possible, leading to different numbers of unknown parameters that need to be estimated, e.g., using an EM-type algorithm; compare \citet{bouveyron_model-based_2011} for details. A Bayesian version of model-based clustering for functional data is, for instance, used in \citet{heard_quantitative_2006}. Extensions to multivariate functional data are, e.g., provided by \citet{jacques_model-based_2014} and \citet{schmutz_clustering_2020}. More generally speaking, all those approaches correspond to the `adaptive methods' from above, and have to be distinguished from the much simpler, two-stage approach of calculating principal component scores on the entire data set first, and then using those values as input to an algorithm for model-based clustering of usual, multivariate data (which would be a `filtering method'). Analogously, distances between functions may be computed/defined via a basis expansion and corresponding basis coefficients, and then be used as input to hierarchical clustering or k-means, which can be described as `dimension reduction in combination with data/observation-based clustering'. By contrast, if computing distances directly from raw data, we would end up with a raw, observation-based functional clustering. A discussion of k-means, dimension reduction (particularly FPCA), and specific cases of equivalence between approaches is found in \citet{Tarpey.2003}. Of course, in addition to dimension reduction, there may also be other pre-processing steps before calculating distances between functions; see, e.g., \citet{ieva_multivariate_2013}.

\begin{figure}[htb]
\begin{center}
\includegraphics[width=70mm]{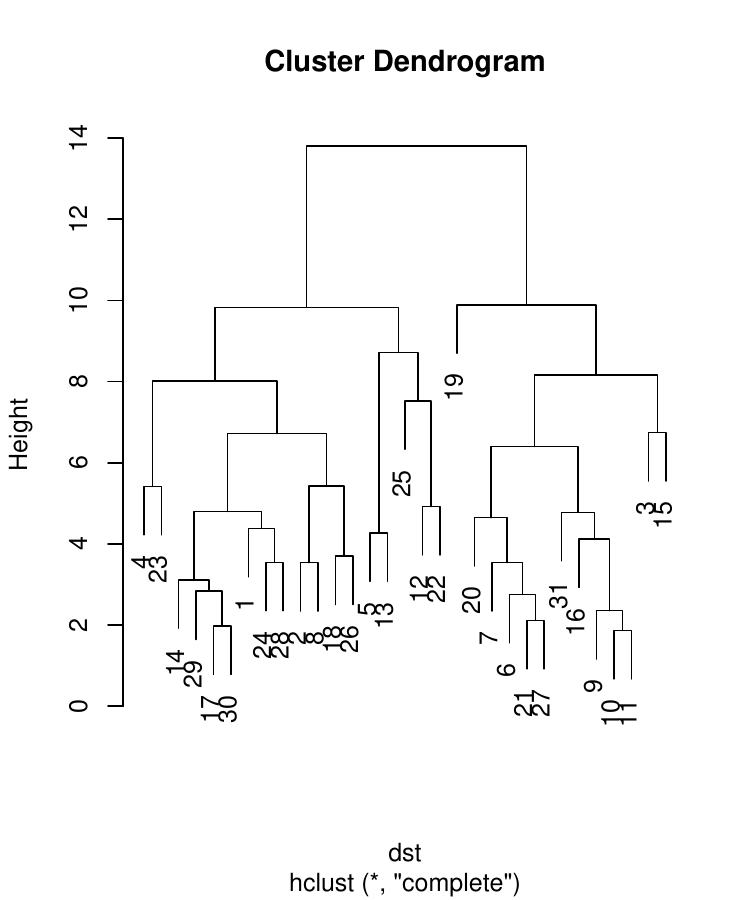}
\includegraphics[width=70mm]{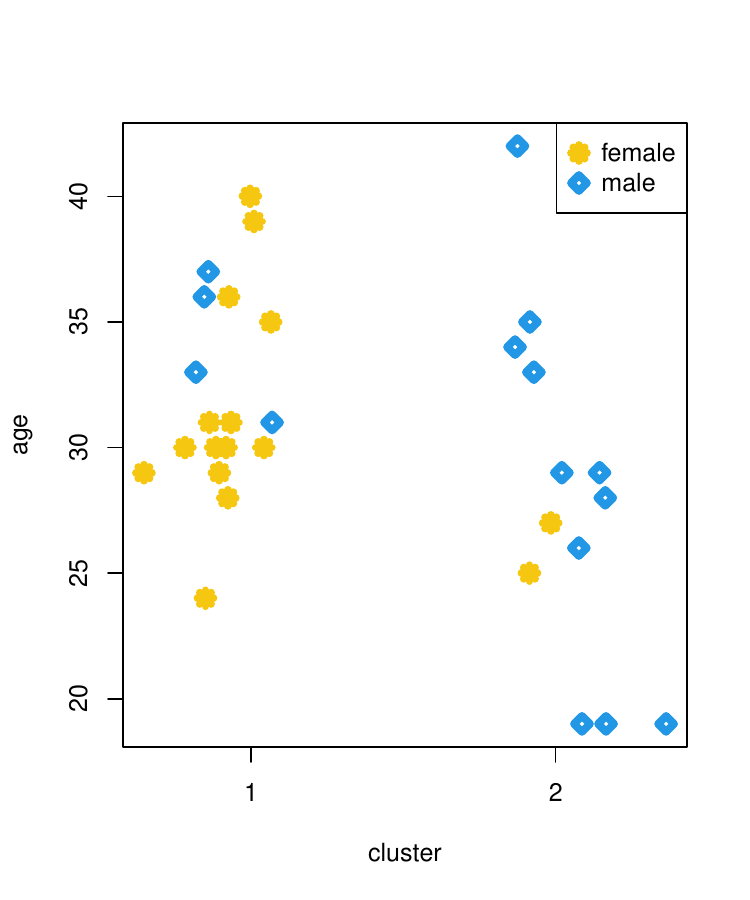}
\caption{Results for hierarchical clustering with complete linkage of functional data from Figure~\ref{fig:examples} (left), with dendrogram (left) and visualization of the 2-cluster solution by age and sex (right).}\label{fig:hclust}
\end{center}
\end{figure}

\begin{figure}[!htb]
\begin{center}
\includegraphics[width=70mm]{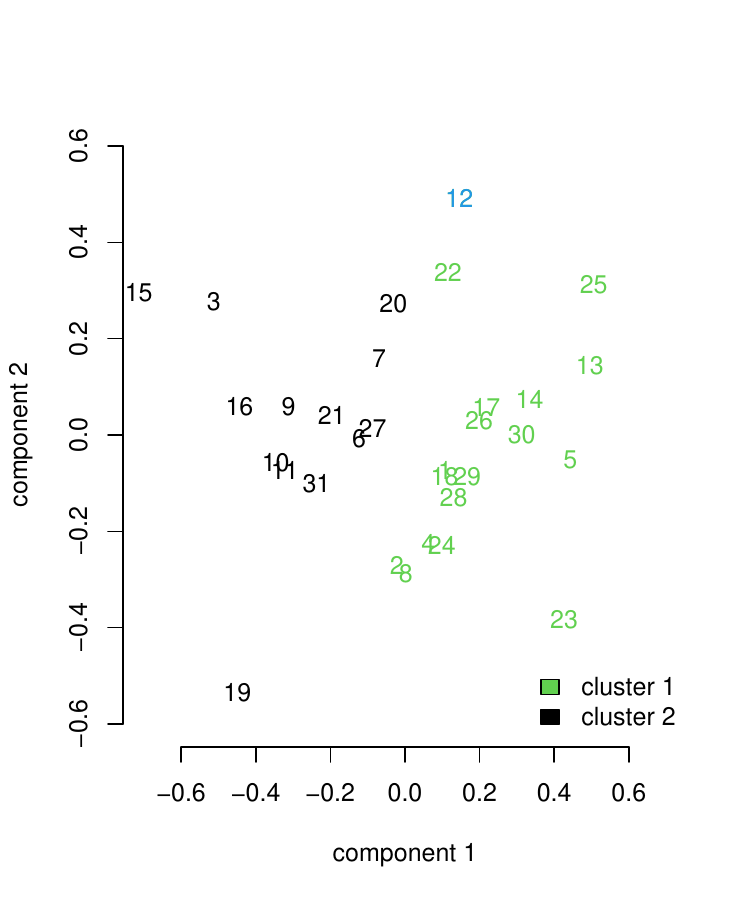}
\includegraphics[width=70mm]{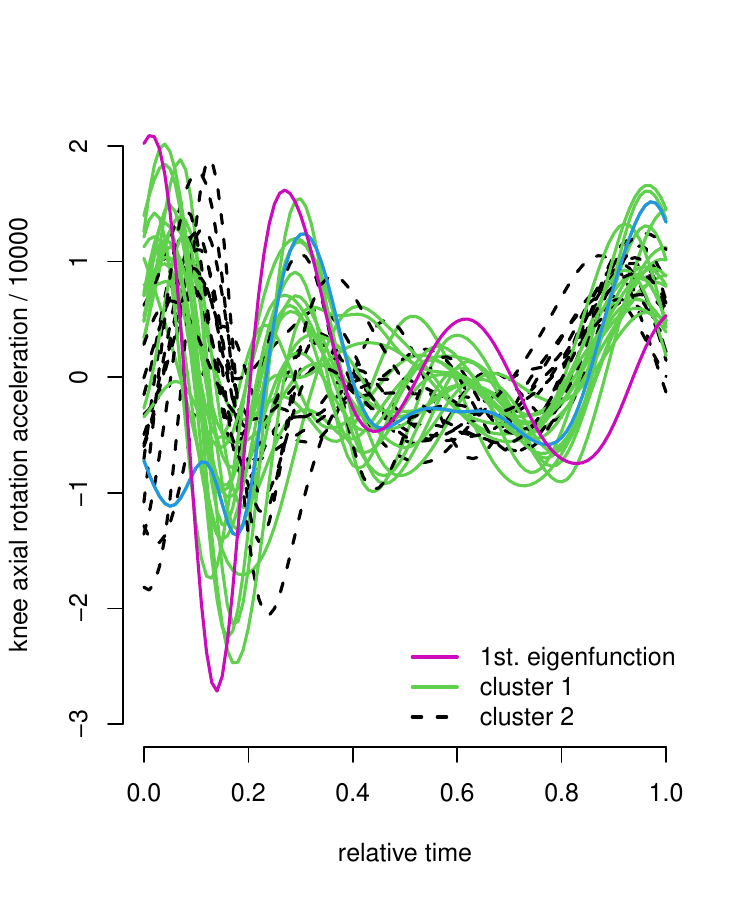}
\caption{Results for k-means clustering using principal component scores of functional data from Figure~\ref{fig:examples}~(left), with clusters in the space of the first two principal components (left) and visualization of the 2-cluster solution in the functions' space together with the first eigenfunction (right). Note, this gives the same result as hierarchical clustering with complete linkage using the raw data (compare Figure~\ref{fig:hclust}, left). The blue color indicates the one curve (12) that moves from cluster 1 into cluster 2 if using model-based, adaptive clustering.}\label{fig:kmeans}
\end{center}
\end{figure}

For illustration, let us consider the curves from Figure~\ref{fig:examples}~(left). To start with, we use the Euclidean distance on the raw data (compare~\ref{subsec:nonp}) as input to hierarchical clustering with complete linkage. The resulting dendrogram is found in Figure~\ref{fig:hclust}~(left), where we can nicely see how curves are successively grouped (bottom-up) into clusters. From the dendrogram, we also get the impression that the data set essentially consists of two clusters. Those two clusters in the space of the first two principal components are found in Figure~\ref{fig:kmeans}~(left). The distribution of age and sex in the clusters is shown in Figure~\ref{fig:hclust}~(right). We see that females are primarily found in cluster 1 whereas cluster 2 mainly consists of males. On average, people in cluster 1 are a little bit older, but the overlap of clusters is much larger for age than sex. Instead of using the raw data, we could also carry out FPCA first. Then, we could use the component scores as input to a standard clustering algorithm for multivariate data, such as hierarchical clustering or k-means. The latter approach minimizes the so-called \emph{within-cluster sum of squares} across all potential partitions into $k$ clusters. With $k=2$, we obtain the clusters as shown in Figure~\ref{fig:kmeans}. If looking at the scores of the first two principal components (Figure~\ref{fig:kmeans}, left), we see that clusters are primarily determined by the first principal component. The first eigenfunction is depicted in pink in Figure~\ref{fig:kmeans}~(right), together with the functional data observed and the clustering. In particular, in the plot's left part, we see how small (i.e., negative) values in the first principal component translate into cluster 2 membership, as cluster 2 mainly consists of curves with a first peak shifted to the right (dashed black). Also, it should be noted that the clusters that are obtained with k-means of FPCA scores (Figure~\ref{fig:kmeans}) are precisely the same as those obtained with hierarchical clustering/complete linkage on the raw data above (Figure~\ref{fig:hclust}).          
A minimal change in the clusters is observed if using the model-based, adaptive approach by \cite{bouveyron_model-based_2011}, where different FPCA solutions are allowed between clusters. Then, one observation (female, age 24), as marked in blue in Figure~\ref{fig:kmeans}, moves from cluster 1 into cluster 2. This curve is somewhat peculiar regarding the beginning of the cycle; compare the blue line in Figure~\ref{fig:kmeans}~(right). Otherwise, however, clustering results are very consistent across different algorithms here.

\subsection{Software}

After dimension reduction (FPCA, spline basis, etc.) using already discussed functions, e.g., from R packages \texttt{fda} or \texttt{refund}, estimated component scores or basis coefficients can simply be used as input to standard cluster algorithms such as \texttt{hclust()} or \texttt{kmeans()} from R's basic \texttt{stats} package~\citep{R_core}; compare Figure~\ref{fig:kmeans}. Also functions from the \texttt{cluster}~\citep{R_cluster} package can be used. If algorithms work on distance/similarity matrices, the latter may also be computed on raw data, e.g., using the \texttt{dist()} function in R, as has been done above (compare Figure~\ref{fig:hclust}). Model-based clustering for multivariate data is, for instance, provided by \citet{R_mclust} through R-package \texttt{mclust}. A functional version as employed above \citep{bouveyron_model-based_2011, schmutz_clustering_2020} is available in \texttt{funHDDC}~\citep{R_funHDDC}.       

\section{Machine Learning Approaches}
\label{sec:ml}

Functional data analysis can also be found in machine learning (ML) and its deep learning (DL) subcategory. We here mainly focus on models for supervised learning. In contrast to previously discussed methods, analyses in ML and DL mainly target predictive performance and the incorporation of functional data primarily serves as an appropriate data representation, rather than to allow for an interpretable relationship of (functional) model in- and outputs. We will also outline a few approaches that deal with unsupervised learning or generative approaches and are related to unsupervised approaches described in Section~\ref{sec:beyond}.

\subsection{Implicit Approaches}

In analogy to the different types of regression (FOSR, SOFR, and FOFR) in previous sections, models in ML are categorized by the nature of their input and output. For a scalar outcome, a distinction is usually made between a \emph{classification} and a \emph{regression task}. Classification includes statistical modeling tasks where the outcome is of discrete nature, e.g., Bernoulli- or multinoulli-distributed, while regression in ML is often used as an umbrella term for all modeling tasks including continuous outcomes (e.g., normally distributed response). Many different ML approaches only implicitly deal with functional data and do not explicitly account for the functional nature of variables, e.g., the smoothness of the inputs. Instead, functional covariates are considered as sequences or (multivariate) time series \citep[see, e.g.,][]{Ott.2021}. Another common practice is to calculate summary statistics of the functional inputs (the mean of the function, location of the mode, different frequency statistics, etc.) to characterize inputs using scalar values instead of using the actual functions \citep[e.g.,][]{vogel.2022}. These statistics can subsequently be used with classical ML approaches intended for scalar data inputs and pre-processing steps turning functional into scalar values are defined as a (tunable) pre-processing pipeline. \citet{Pfisterer.2019} compare such classical ML approaches with functional classification algorithms described in Section~\ref{sec:sofr}. An alternative representation of functional inputs in, e.g., the field of biomechanics, is a picture representation of multivariate functions \citep{liew.2021}. Given all functional inputs have the same domain $\mathcal{T}$, the approach concatenates the $p$ functional observations $(x_{i1}(t_v),\ldots,x_{ip}(t_v))$ for $v=1,\ldots,V$ to a $V \times p$ dimensional matrix for every observation $i$, which is then considered as an ``image'' in the subsequent analysis. This allows the use of pre-trained models and is based on the idea of transfer learning \citep[see, e.g.,][]{Pan.2010}. For functional outcomes, ML methods are often referred to as trajectory regression \citep{Ott.2021} or sequence-to-sequence models \citep{Sutskever.2014}. Most approaches do not impose smoothness of the output function, but either adjust the loss function to favor smooth(er) outputs or impose regularization to foster generalization and thereby implicitly generate smoother output functions.

In addition to the previously described implicit approaches, there is also a body of literature in ML that explicitly takes the functional nature of objects into account. We describe several methods in the next subsection.

\subsection{Explicit Approaches}

\paragraph{Machine Learning} Among the classic ML approaches, \citet{Rossi.2006} are one of the first to propose an explicit approach for functional data by adapting support vector machines (SVMs) for FD. SVMs are kernel-based methods, but in contrast to the use of kernels in Section~\ref{subsec:nonp}, SVMs utilize kernels $K(\cdot, \cdot)$ as a distance measure between two observations. \citet{Rossi.2006} argue that, although kernels in principle also work for functions in a Hilbert space $\mathcal{H}$, it is difficult to implement calculations in practice and these kernels do not take advantage of the functional nature of the data. The authors derive a simple solution by defining a map $\mathcal{P}$ from $\mathcal{H}$ to the Euclidean space. In this space classic kernels $K(\cdot, \cdot)$ such as the Gaussian or exponential kernel can be straightforwardly implemented and then used as $K(\mathcal{P}(f), \mathcal{P}(g))$ to evaluate the distance between to functions $f,g \in \mathcal{H}$. Another kernel-based ML approach is Gaussian processes (GPs). An adaptation of GPs called Gaussian process functional regression (GPFR) for functional data has been proposed in \citet{shi2011gaussian} with implementation in \citet{Konzen.2021}. GPFR additively combines the mean estimation of a functional regression model with a multivariate GP. The linear (functional) effects of scalar or functional covariates thereby define the mean of the functional response while functional covariate(s) over $\mathcal{T}$ or the time information is used for the estimation of the covariance using the GP. Another important class of ML models is boosting. Gradient boosting approaches for functional regression models and various extensions have been proposed by \citet{Brockhaus.2015, Brockhaus.2017, Ruegamer.2018}. By not relying on regression trees but on single additive predictors, the corresponding boosting implementation \citep{Brockhaus.2020} can be seen as a special coordinate-wise gradient descent optimization routine. The Gauss-Southwell-type update rule in this case guarantees good convergence and additionally induces sparsity due to the coordinate-wise nature of the optimization. This, in particular, allows fitting functional regression models in very high dimensions as the optimization routine only optimizes the effect of one (functional) feature at a time. Other extensions of this approach have been proposed recently, including densities-on-scalar regression models using a Bayes Hilbert space representation \citep{maier2021additive} or functional additive models on manifolds of planar shape and form \citep{stocker2021functional}. Finally, tree-based methods have also been brought forward, both for functional covariates using random forests \citep{Moeller.2016, Rahman.2019}, and also via functional random forests for functional outcomes \citep{fu2021functional}. 

In addition to these supervised learning approaches, several unsupervised methods with explicit function representation exist. The most prominent example is functional data clustering \citep{Tarpey.2003} as explained in detail in Section~\ref{subsec:clust}. An overview can be found in \citet{Jacques.2014}. In DL, one prominent unsupervised learning technique is autoencoders \citep[AEs;][]{Kingma.2013}, which try to encode a high-dimensional input into a small number of hidden variables and then decode these variables back to the original input size with the goal to recover the original input. A recent adaption of AE is given by \citet{Hsieh.2021}, proposing a functional autoencoder. The proposed approach can be seen as a non-linear extension of FPCA. Another method that can be seen as an unsupervised technique from an ML perspective is function registration, which has recently been combined with deep neural networks \citep{chen2021srvfregnet}. The following paragraph will explain the idea of functions in neural networks in some more detail.

\paragraph{Deep Learning and Neural Networks}

\citet{Rossi.2002} are among the first to combine neural networks with functional data by proposing a multi-layer perceptron (MLP) for functional data and also extending the universal approximation theorem of neural networks \citep{Hornik.1989} to functional inputs. An MLP for scalar values is defined by multiple neurons gathered in layers that are in turn stacked onto each other. Every neuron (or unit) can be seen as a function $h: \mathbb{R} \to \mathbb{R}, x \mapsto \tau(b + wx)$ that takes a scalar value $x$ and computes a hidden state value of the form $\tau(b + wx)$ with bias (intercept) $b$, weight (coefficient) $w$ and activation (response) function $\tau$. Similar to scalar-on-function regression, a functional analog for functional inputs $x$ can be defined by $\tau(b + \int w(t)x(t)dt)$ and approximated at given measurement points by finding a suitable approximation for $w(t)$ and the integral over $t$. This will result in units $h: \mathcal{H} \to \mathbb{R}$ that take functions as input and output a scalar value. Thus, the MLP is only functional in its first layer, while the following layers consist of conventional layers and neurons. In recent years, various other authors have proposed extensions and unifications \citep{Guss.2016, Wang.2019, Thind.2020, Rao.2021}. For functional response variables, the functional MLP can be extended in an analogous manner to the extension from a scalar-on-function to a function-on-function regression. The $k$th output neuron $h_k^{(l)}$ of a functional layer $l=1,\ldots,L$ in a functional MLP with $L$ layers, each with $J_l$ output neurons, is recursively defined by its previous layers as 
\begin{equation}
    h^{(l)}_k(t) = \tau^{(l)}\left(b_k^{(l)}(t) + \sum_{j=1}^{J_{l-1}} \int w_{j,k}^{(l)}(s,t) h^{(l-1)}_j(s) ds \right), 
\end{equation}
where $b_k^{(l)} \in \mathcal{L}^2(\mathcal{T})$ and $w_{j,k}^{(l)} \in \mathcal{L}^2(\mathcal{T} \times \mathcal{T})$. In addition, special roles are given to the input layer by defining $h_j^{(0)}(\cdot) = x_j(\cdot)$ for all covariates $j=1,\ldots,p$, and the very last layer consisting of only one unit $J_L = 1$ defined as $h^{(L)}(t) = \mathbb{E}(Y(t) | \boldsymbol{X})$, where $\boldsymbol{X}$ collects all $p$ functional covariates. Similar to the mean function, it is also possible to learn the covariance kernel using deep neural networks \citep{Sarkar.2021}. An extension to neural networks that model functional data both with structured predictors and arbitrary deep neural network architectures has been recently proposed by \citet{ruegamer2024}.

\subsection{Benchmark Comparison}

\begin{figure}[!ht]
\begin{center}
\includegraphics[width=\textwidth]{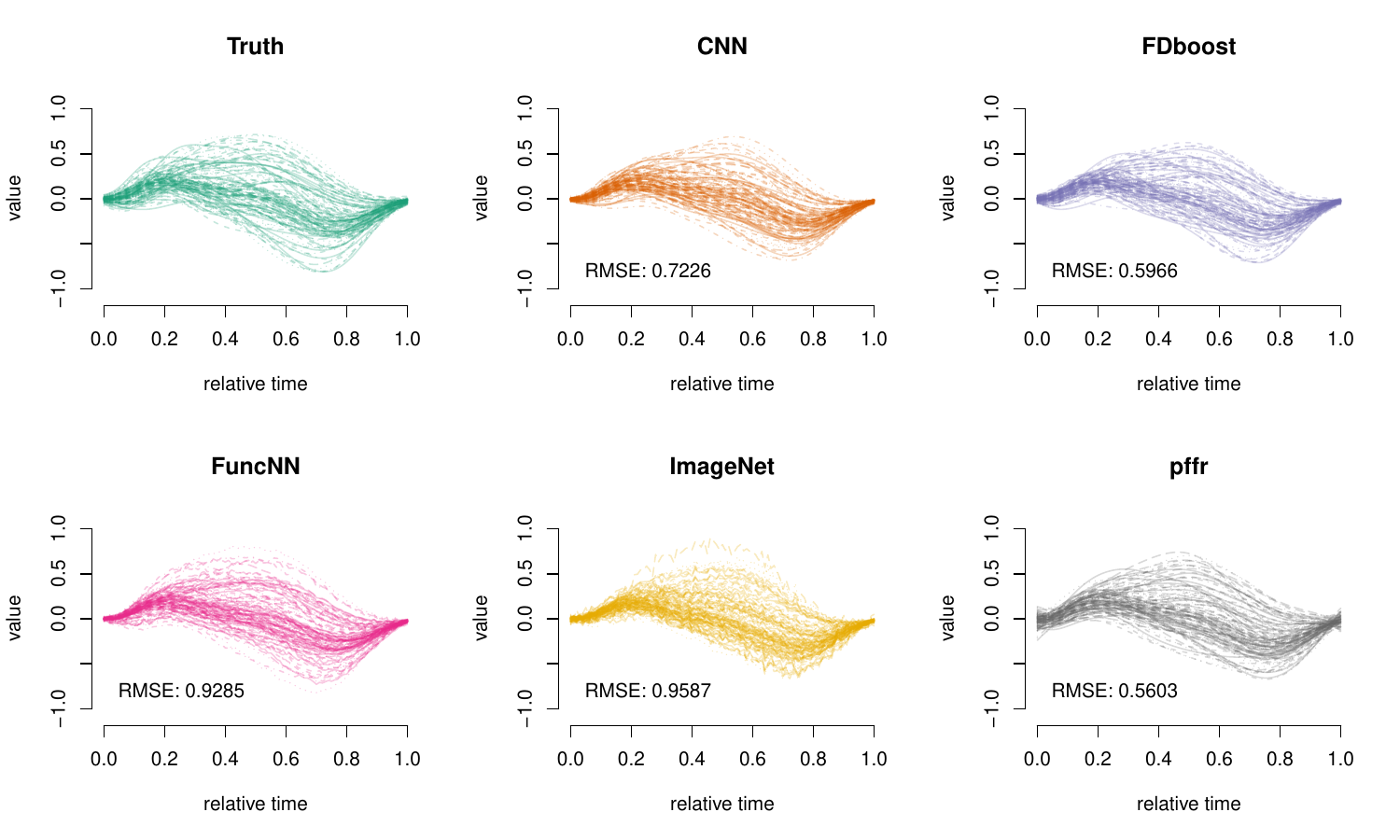}
\caption{Comparison of different ML and DL methods (facets) to fit a function-on-function relationship on the running data with respective test RMSE values for every method in the bottom left corner.}\label{fig:ml}
\end{center}
\end{figure}

To demonstrate the performance of different methods presented in this section, we conduct a small benchmark comparison between selected methods with available software implementations. 
We want to emphasize that all ML and DL methods are not tuned but either used with default settings or trained with values from preliminary experiments. We use two implicit neural network approaches as proposed in \cite{liew.2021} that represent functional covariates as images. One network is a classic convolutional neural network (CNN) with several convolution blocks to process the ``image'' and predict a vector of the length of the discretized output function $y(t)$. The second implicit neural network approach uses a pre-trained deep neural network (more specifically the ImageNet architecture \citep{imagenet_cvpr09}) and applies transfer learning for the given ``images''. We also use an explicit functional neural network as proposed by \cite{Thind.2020} (FuncNN), the boosting approach FDboost by \cite{Brockhaus.2020} and compare these approaches against a statistical approach, the \texttt{pffr} function for penalized function-on-function regression in \cite{R_refund}, see also Section~\ref{sec:funreg}. We split the data into a train and test set, using the ankle moment (inversion-eversion) as the outcome function and the different joint accelerations, velocities, and angles of all three body parts as functional covariates. For all methods except for the CNNs, we additionally use the scalar covariates age, height, weight, and sex as scalar features. For structured models (\texttt{pffr}, \texttt{FDboost}), we also incorporate random effects for the running condition and the study. Figure~\ref{fig:ml} depicts the true functions in the test set as well as predictions of the five methods. In addition, test performance results for the average point-wise root mean squared error (RMSE) between actual and predicted functions is given for every method. Results suggest that methods all work well in general capturing the overall trend in the data, but the approaches estimating an additive model (\texttt{FDboost}, \texttt{pffr}) perform better in the details than the different neural network approaches.  

\subsection{Software}
Software packages that model functional data explicitly using machine and deep learning include \texttt{FDboost} \citep{Brockhaus.2017} building on a model-based boosting framework, \texttt{FuncNN} \citep{Thind.2020} for functional neural networks, \texttt{funGp} \citep{funGp} using GPs for functional inputs or \texttt{GPFDA} \citep{gpfda} to combine functional regression with GPs. 

\section{Outlook} \label{sec:outlook}
\subsection{Functional data and related fields}

While we discussed the differences between FDA, LDA, and time series analysis in the introduction, developments in functional data analysis have also led to fruitful cross-fertilization in these other two areas. Depending on the goal of the analysis and the data structure, especially the semi/nonparametric analysis methods
of FDA can also be of interest in these two fields. In particular, 
longitudinal data can be viewed as sparse (and noisy) functional data. 
FDA then provides new perspectives: the commonly used linear mixed models for longitudinal data are parametric, with, e.g., a random intercept random slope model being linear over time per subject (observational unit). On the other hand, FPCA with its subject-specific random scores for the general principal component (PC) functions, can be viewed as a linear mixed model with data-driven PC basis functions replacing the constant and linear basis functions, yielding larger flexibility and a better semi-parametric fit to the longitudinal data \citep[e.g.][]{yao2005functional,goldsmith2013corrected}. 
Similar approaches have also been successfully incorporated  in 
joint models for longitudinal and time-to-event data to allow for more flexible modeling of the univariate or multivariate longitudinal trajectories \citep[e.g.][]{kohler2017flexible,kohler2018nonlinear,li2022joint,volkmannetal2023}. 
While time series analysis usually focuses on one time series, there are also cases where a sample of time series is collected and questions such as time series classification are of interest, in which case functional data classification methods can be utilized as discussed in Sections \ref{subsec:class} and \ref{sec:ml}. Generally, there are often 
different ways to look at the same data depending on the research questions of interest. In addition to longitudinal and time series data, this also holds, e.g., for spatio-temporal data, which depending on the setting and questions of interest could also often be viewed as repeated observations of spatial random fields over time, or spatially correlated curves  (time series or functional data) over time \citep[e.g.,][]{delicado2010statistics,kokoszka2012dependent}. Besides spatial correlation, functional data can appear with other additional structures known from scalar data, such as, e.g., functional time series  \cite[e.g.,][]{hyndman2009forecasting,kokoszka2012dependent,aue2015prediction} or longitudinal functional data \cite[e.g.,][]{greven2011longitudinal,park2015longitudinal}. 
\citet{KonSta2023} call multivariate functional data, longitudinal functional data, functional time series, and spatially correlated functional data ``second-generation functional data'' and provide a review.

\subsection{Beyond one-dimensional functions}

Functional data analysis is still an active field of research with many current developments. In recent years, an important focus of the field has been to broaden the view and see functional data as just one example of more general `object data', i.e., of data where each observation constitutes an object in a more complex space than the usual $\mathbb{R}^d$ space for vectors. This area is sometimes called object (oriented) data analysis \citep{marron2014overview}. While the univariate functional data considered here are often viewed as objects in the $\mathcal{L}^2(\mathcal{T})$ of square-integrable functions on some interval $\mathcal{T}\subset \mathbb{R}$, images and surface-valued data \citep[e.g.,][]{goldsmith_smooth_2014,happ2018multivariate} can be viewed as $\mathcal{L}^2(\mathcal{T})$ functions on a higher-dimensional domain $\mathcal{T}\subset \mathbb{R}^d$, $d >1$, while multivariate functions are often defined on an interval $\mathcal{T}$ but to be vector-valued in $\mathbb{R}^d$, $d >1$ \citep[e.g.,][]{chiou2014multivariate,gorecki2018selected}, with a more general setting considered in \citep{happ2018multivariate}. Examples are brain scans for the former, and movement trajectories, outlines, or several recorded functions for the latter \citep[e.g.,][]{steyer2022elastic,steyer2023elastic,volkmann2023multivariate}. Many methods developed for univariate functional data can be suitably extended to images and multivariate functional data such as, e.g., functional principal component analysis \citep[e.g.,][]{berrendero2011principal,chiou2014multivariate,happ2018multivariate}  or regression models \citep[e.g.,][]{goldsmith_smooth_2014,chiou2016multivariate,volkmann2023multivariate}. For both settings, phase variation can also be important, for image analysis due to necessary registration. For multivariate functional data, it can also be relevant to have the analysis be invariant to warping, i.e., only analyze the image of a curve, such as when an outline of an object such as a cell or bone is recorded and the parametrization of the curve along that outline is essentially arbitrary. In this case, the object of interest is the equivalence class of such curves with respect to reparametrization, and methods such as distance and mean computation as well as regression have also been developed for such elements of quotient spaces \cite[e.g.,][]{srivastava2010shape,srivastava2016functional,steyer2022elastic,steyer2023elastic}. Similarly, the shape/form of an object (an outline or vector of landmark coordinates) is considered to be the equivalence class under rotation, translation, and (for shape) scale \citep{dryden2016statistical}, and is of interest if the coordinate system of the recorded object is arbitrary or not relevant. Such shapes/forms live on certain manifolds \citep[e.g.,][]{huckemann2009intrinsic,thomas2013geodesic,stocker2021functional} or in more complex quotient spaces, e.g., if parametrization invariance is simultaneously considered \citep[e.g.][]{stocker2022elastic,steyer2022elastic}.
Shape analysis \citep{dryden2016statistical} is an important field in itself and has more recently intersected with functional data analysis in functional shape analysis \citep{srivastava2016functional} when the shape of a curve is of interest, e.g., as the outcome in a regression model \citep{stocker2021functional}. 
Similarly, compositional data analysis 
\citep{CoDawithR,PawlowskyGlahn2015,Filzmoser2018} is a field with a long tradition, looking at multivariate vectors of non-negative entries that add up to a constant such as 1 or 100\%. It takes into account the special geometry of the simplex, which compositions are elements of. More recently, this field has also intersected with functional data analysis when looking at probability densities, which can be viewed as functional data with particular constraints or as infinite compositions. More generally, continuous, discrete, mixed, bivariate, etc.\ densities can be considered \citep[e.g.,][]{maier2021additive}, with one possible geometry given by the so-called Bayes-Hilbert-space approach \citep{vdb2010,vdb2014} 
and compositions seen as the special case of discrete densities. Other possible approaches include Wasserstein spaces \citep{panaretos2020invitation}. Further examples of object data include so-called generalized functional data \citep[e.g.,][]{HalMueYao:2008,goldsmith2015generalized,scheipl2016generalized,greven2017general,GerGolSta:2017}, where conditional on a smooth mean curve over $\mathcal{T}$ as in functional data, observed values are viewed as realizations from some non-Gaussian (e.g., binary or count) distribution, directional data \citep{mardia1975statistics,mardia1988directional}, covariance-(matrix- or operator-)valued data \citep[e.g.,][]{masarotto2019procrustes,lin2023additive},  point-process-valued data \citep[e.g.,][]{panaretos2016amplitude}, and tree- or network-valued data  \citep[e.g.,][]{feragen2013tree,duncan2018statistical,calissano2022graph,calissano2023populations}. In contrast to neighboring fields such as graphical models, the focus is not on modeling, e.g., a single network, but always on the setting where a sample of networks or more generally objects is observed. The objects constitute the unit of analysis, e.g., the response or covariate in a regression model. Many developments are expected to occur in this exciting field in the coming years.

\section*{Acknowledgement}
Sonja Greven gratefully acknowledges funding by grants GR 3793/2-2, 
GR 3793/3-1 
and GR 3793/8-1 
from the German research foundation (DFG). Bernard Liew is supported by The Academy of Medical Sciences, UK, Springboard Award SBF006/1019.
\vspace*{1pc}

\bibliographystyle{abbrvnat}
\bibliography{bibliography.bib}
\end{document}